\documentclass[%
 reprint,
 amsmath,amssymb,
 aps,
]{revtex4-1}

\usepackage{graphicx}
\usepackage{dcolumn}
\usepackage{bm}
\usepackage{color}
\usepackage{ulem}

\newcommand{\simgt}{\lower.5ex\hbox{$\; \buildrel > \over \sim \;$}}
\newcommand{\simlt}{\lower.5ex\hbox{$\; \buildrel < \over \sim \;$}}

\begin{document}

\preprint{APS/123-QED}

\title{Covariance of the redshift-space matter power spectrum after reconstruction}

\author{Chiaki Hikage}
\affiliation{%
 Kavli Institute for the Physics and Mathematics of the Universe (Kavli IPMU, WPI), University of Tokyo, 5-1-5 Kashiwanoha, Kashiwa, Chiba, 277-8583, Japan
}%
 \email{chiaki.hikage@ipmu.jp}
\author{Ryuichi Takahashi}%
\affiliation{%
 Faculty of Science and Technology, Hirosaki University, 3 Bunkyo-cho, Hirosaki, Aomori 036-8588, Japan
 }%
\author{Kazuya Koyama}%
\affiliation{%
 Institute of Cosmology and Gravitation, University of Portsmouth, Portsmouth PO1 3FX, UK
}%

\date{\today}

\begin{abstract}
We explore the covariance of redshift-space matter power spectra after a standard density-field reconstruction. We derive perturbative formula of the covariance at the tree-level order and find that the amplitude of the off-diagonal components from the trispectrum decreases by reconstruction. Using a large set of $N$-body simulations, we also find the similar reduction of the off-diagonal components of the covariance and thereby the signal-to-noise ratio (S/N) of the post- reconstructed (post-rec) power spectra significantly increases compared to the pre-reconstructed (pre-rec) spectra. This indicates that the information leaking to higher-order statistics come back to the two-point statistics by reconstruction. Interestingly, the post-rec spectra have higher S/N than the linear spectrum with Gaussian covariance when the scale of reconstruction characterized with the smoothing scale of the shift field is below $\sim 10h^{-1}$Mpc where the trispectrum becomes negative. We demonstrate that the error of the growth rate estimated from the monopole and quadrupole components of the redshift-space matter power spectra significantly improves by reconstruction. We also find a similar improvement of the growth rate even when taking into account the super-sample covariance, while the reconstruction cannot correct for the field variation of the super-sample modes.
\end{abstract}
\maketitle

\section{Introduction}

A biggest challenge in the modern cosmology is the mystery of dark matter and dark energy \citep[e.g.,][]{Amendola05}. Large-scale structure traced by galaxies is one of powerful cosmological probes to study the properties of the dark components. The baryonic acoustic oscillation (BAO) imprinted on galaxy distributions is a powerful cosmological probe to study the expansion history of the Universe \citep{Eisenstein98,Meiksin99,BlakeGlazebrook03,HuHaiman03,Matsubara04,Angulo05,SeoEisenstein05,White05,Eisenstein05,Cole05,Padmanabhan07,Percival07,Eisenstein07a,Huff07,Angulo08,Padmanabhan07,Percival07,Okumura08,Xu13,Anderson14,Tojeiro14,Kazin14,Ross15,Alam16,Beutler17}. Bulk motion of galaxies associated with the growth of the large-scale structure generates the anisotropy in the redshift-space galaxy distribution, i.e., the redshift-space distortion, which probes the growth rate of the large-scale structure and is useful to test General Relativity and modified gravity models \citep[e.g.,][]{Guzzo08,Yamamoto08,Reid12,Beutler13,Samushia13,Oka13,HY2013}. The full shape of the power spectrum also has fruitful information of cosmology \citep[e.g.,][]{Ivanov20}. Upcoming spectroscopic galaxy surveys such as PFS \citep{PFS}, DESI \citep{DESI16}, HETDEX \citep{HETDEX}, Euclid \citep{Euclid16} and WFIRST\citep{WFIRST15} are expected to do precise cosmological studies to clarify the nature of dark matter and dark energy. 

In the linear perturbation theory, different wavelength modes of the fluctuations of matter density field grow independently. Since the gravitational growth of the large-scale structure is a nonlinear process, different modes are coupled with each other, which makes a precise cosmological analysis difficult. For example, the BAO signal in the large-scale structure is degraded and the BAO scale is biased at later times \citep[e.g.,][]{CrocceScoccimarro08}. The perturbation theory breaks down in the nonlinear regime and thus precise analytical prediction is difficult at small scale. The two-point statistics such as the power spectrum characterize the whole statistical properties in Gaussian fields. However, non-Gaussianity increases in the evolved matter density field and thereby higher order moments/correlation functions also become important. The information content of the power spectrum are shown to be saturated on nonlinear scale \citep{Scoccimarro99b,Takahashi09,Carron15}. This indicates that the cosmological information leaks to higher-order statistics beyond two-point statistics, which makes our cosmological analysis more complicated. 

Density-field reconstruction aims for recovering the initial or linearly evolved density field. A standard BAO reconstruction method proposed by \citep{Eisenstein07b} shifts mass particles or galaxies toward their initial Lagrangian positions to recover the original BAO signal. The shift field is estimated using the inverse Zeldovich approximation \citep{Zeldovich70} from the observed (evolved) density field after smoothing small-scale power. The reconstruction effectively undoes the bulk motion and thereby the BAO signal is successfully recovered \citep{Seo08,Padmanabhan09,Noh09,Seo10,Sherwin12}. The BAO reconstruction method has been applied to various galaxy surveys \citep{Padmanabhan12,Xu13,Anderson14,Tojeiro14,Kazin14,Ross15,Alam16,Beutler17}. It was also shown that the correlation of the reconstructed matter density field with the initial density field extends to more nonlinear scale \citep{Seo10,TassevZaldarriaga12,Schmittfull15,Seo16,Schmittfull17,Wang17,Yu17,Hada18}. Ref.~\cite{HKH17} derived the 1-loop perturbative formula of the reconstructed matter power spectrum in real space. They found that the amplitudes of the 1-loop nonlinear terms decrease substantially by reconstruction and the perturbation theory works at more nonlinear scale. Ref.~\citep{HKT19} extended their formula to redshift-space clustering and demonstrated that the growth rate measurement from the redshift-space distortion is significantly improved. 

How does the reconstruction alter the covariance of the matter power spectra? The covariance of matter power spectra without reconstruction has been investigated both from the perturbation theory and from $N$-body simulations. The tree-level perturbative formula of the covariance of the matter power spectra was derived to show that the non-Gaussian effect generates the correlations between different bands and thereby the covariance has non-zero off-diagonal components \citep[e.g.,][]{Scoccimarro99b,Meiksin99}. The matter covariance has been also investigated from a large set of N-body simulations to show that the non-Gaussian effects significantly suppresses the signal-to-noise ratio (S/N) of the power spectrum and thereby degrades the information content of the power spectrum \citep{Takahashi09,HarnoisDeraps13,Blot15,Klypin18,Quijote19}. The non-Gaussian effects on the matter covariance comes from the mode coupling associated with the nonlinear gravity. Since the reconstruction effectively linearizes the field by partially removing mode-coupling effect, it is expected that the covariance of the power spectrum is more diagonalized.

Here we first investigate the covariance of the redshift-space matter power spectra after the field is reconstructed and evaluate the information content of the matter power spectrum by using the perturbation theory and a large set of N-body simulations. The mode coupling between the small-scale modes and the large-scale modes beyond survey size has a significant contribution to the matter covariance, which is known as `beat coupling' \citep{Hamilton06} or `super-sample covariance (SSC)' \citep{TakadaHu13,Li14}. We also investigate the effect of SSC on the reconstructed covariance. Finally, we show the improvement of the growth rate measurement from the redshift-space power spectra when using the covariance matrix of reconstructed spectra.

This paper is organized as follows: in Section \ref{sec:perturbation}, we derive a tree-level perturbative formula of the covariance of monopole and quadrupole components of matter power spectra after reconstructing the field. We see how the off-diagonal components of their covariance are changed by reconstruction with different smoothing scales. In Section \ref{sec:nbody}, we also numerically estimate the covariance using a large set of $N$-body simulations to see the behavior of the covariance of reconstructed spectra. In Section \ref{sec:SN}, we evaluate the S/N of the reconstructed power spectra to discuss how much the information content is recovered. In Section \ref{sec:growthrate}, we study the impact on the growth rate measurement by using the covariance of reconstructed spectra. We also study the effect of the super-sample covariance on our results in Section \ref{sec:ssc}. Section \ref{sec:summary} is devoted to the summary and conclusions. 

Throughout this paper, we assume a flat $\Lambda$CDM model with the best-fit values of Planck TT,TE,EE+lowP in 2015, i.e.,
$\Omega_b=0.0492$, $\Omega_m=0.3156$, $h=0.6727$, $n_s=0.9645$, and $\sigma_8=0.831$ \citep{Planck15}. 

\section{Perturbative formula of covariance of matter power spectra in redshift space}
\label{sec:perturbation}

In this section, we derive the covariance of the monopole ($\ell=0$) and quadrupole ($\ell=2$) components of the redshift-space matter power spectrum in a perturbative approach. Our formula is applicable to higher-order multipoles such as hexadecapole ($\ell=4$), however, we do not include them because their signal-to-noise ratios are small relative to the monopole and the quadrupole and also their signals are dominated by the nonlinear redshift-space distortion.

The covariance can be generally decomposed into the Gaussian and the non-Gaussian parts as
\begin{equation}
\mathbf{Cov}=\mathbf{Cov}^{\rm (G)}+\mathbf{Cov}^{\rm (NG)}.
\end{equation}
When neglecting the convolution with the survey geometry, the Gaussian part is given by

\begin{eqnarray}
\label{eq:cov_ga}
\mathbf{Cov}_{\ell\ell^\prime}^{\rm (G)}(k_i,k_j)&=&
\frac{(2\ell+1)(2\ell^\prime+1)}{2}\int_{-1}^1d\mu
{\cal L}_\ell(\mu_{\mathbf{k}_i}){\cal L}_{\ell^\prime}(\mu_{\mathbf{k}_j}) \nonumber \\
&& \frac{2}{V}\frac{(2\pi)^3}{V_{k_i}}\delta_{ij}^{\rm K}\left[P^{\rm z}(\mathbf{k}_i)+\frac{1}{n}\right]\left[P^{\rm z}(\mathbf{k}_j)+\frac{1}{n}\right], \nonumber \\
\end{eqnarray}
where $\mu_{\mathbf{k}}$ is the cosine angle between $\mathbf{k}$ and the line-of-sight direction, $\cal{L}_\ell$ is the $\ell$-th Legendre polynomial, i.e., ${\cal L}_0=1$ and ${\cal L}_2=(3\mu^2-1)/2$, $V$ is the sample volume, and $n$ is the number density of mass particles. The volume of $k$-binning shell $V_{k_i}$ is approximated as $4\pi k_i^2\Delta k$ where the binning width $\Delta k$ is much smaller the mean wavenumber of $i$-th bin $k_i$.
The redshift-space linear matter power spectrum $P^{\rm z}(\mathbf{k})$ is given by
\begin{equation}
P^{\rm z}(\mathbf{k})=Z_1(\mathbf{k})^2P_{\rm L}(k),
\end{equation}
where $P_{\rm L}(k)$ is the linear matter power spectrum in real space and $Z_1$ is the first-order Eulerian kernel in redshift space
\begin{equation}
Z_1(\mathbf{k})=1+f\mu_{\mathbf k}^2,
\end{equation}
with the linear growth rate $f\equiv d\ln D/d\ln a$ defined as the logarithmic derivative of the linear growth factor. 

\begin{figure*}
\begin{center}
\includegraphics[width=18cm]{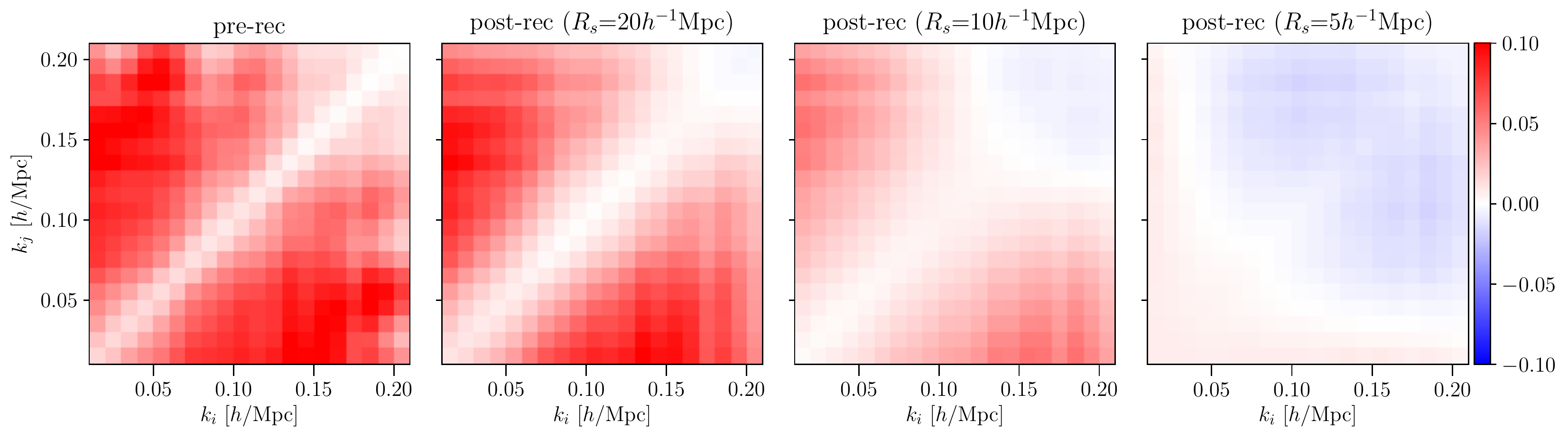}
\caption{Tree-level non-Gaussian covariance of the monopole component of the matter power spectrum normalized with their Gaussian covariance, i.e., ${\bf Cov}_{00}^{\rm (tree)}(k_i,k_j)/[{\bf Cov}_{00}^{\rm (G)}(k_i,k_i){\bf Cov}_{00}^{\rm (G)}(k_j,k_j)]^{1/2}$ for pre-rec and post-rec spectra with $R_s=20h^{-1}$Mpc, $10h^{-1}$Mpc and 5$h^{-1}$Mpc from left to right panels. As $R_s$ is smaller, the off-diagonal terms are changed from positive to negative values. The output redshift is $z=1.02$.}
\label{fig:cov_tree}
\end{center}
\end{figure*}
Next we derive the non-Gaussian covariance at tree level. We do not take into account the higher-order covariance such as the one-loop covariance for simplicity. The one-loop covariance becomes important at higher $k$, however, the shot noise usually dominates the error at higher $k$ in actual observed data. In section \ref{sec:ssc}, we take into account the super-sample covariance (SSC).

The tree-level covariance comes from the tree-level term of the trispectrum of matter power spectra given by \citep{Fry84}
\begin{eqnarray}
T^{\rm (tree)}(\mathbf{k_1},&\mathbf{k_2}&,\mathbf{k_3},\mathbf{k_4})=
4[Z_2(\mathbf{k_{12}},-\mathbf{k_1})
Z_2(\mathbf{k_{12}},\mathbf{k_3})P_{\rm L}(k_{12})\nonumber \\
&& \times Z_1(k_1)P_{\rm L}(k_1) Z_1(k_3) P_{\rm L}(k_3)+(11~{\rm perms.})] \nonumber \\
&&+6[Z_3(\mathbf{k_1},\mathbf{k_2},\mathbf{k_3}) Z_1(k_1)P_{\rm L}(k_1) \nonumber \\
&& \times  Z_1(k_2)P_{\rm L}(k_2) Z_1(k_3))P_{\rm L}(k_3)+(3~{\rm perms.})],
\end{eqnarray}
where $Z_n$ is the $n$-th order Eulerian perturbation kernel of the matter density field in the redshift space \citep{Scoccimarro99a,Matsubara08a}.
The tree-level covariance of the multipole power spectra is written as \citep{Scoccimarro99b,Wadekar19}
\begin{eqnarray}
\label{eq:cov_tree}
\mathbf{Cov}_{\ell\ell^\prime}^{\rm (tree)}(&k_i&,k_j)=\frac{1}{V}\int_{\mathbf{\hat{k}}_{i\ell}}\int_{\mathbf{\hat{k}^\prime}_{j\ell^\prime}} 
T^{\rm (tree)}(\mathbf{k},\mathbf{-k},\mathbf{k^\prime},\mathbf{-k^\prime})
\nonumber \\
&& =\frac{1}{V}\int_{\mathbf{\hat{k}}_{i\ell}}\int_{\mathbf{\hat{k}^\prime}_{j\ell^\prime}}  \nonumber \\
&& [12Z_3(\mathbf{k},-\mathbf{k},\mathbf{k^\prime}) Z_1(k)^2 Z_1(k^\prime)P_L(k)^2P_L(k^\prime) \nonumber \\
&& +8Z_2(\mathbf{k-k^\prime},\mathbf{k^\prime})^2 Z_1(k^\prime)^2
P_L(k^\prime)^2P_L(|\mathbf{k^\prime-k}|) \nonumber \\
&& +8Z_2(\mathbf{k-k^\prime},\mathbf{k^\prime})
Z_2(\mathbf{k^\prime-k},\mathbf{k})P_L(|\mathbf{k^\prime-k}|)  \nonumber \\
&& \times  Z_1(k)P_L(k) Z_1(k^\prime)P_L(k^\prime)
+(\mathbf{k}\leftrightarrow \mathbf{k^\prime})],
\end{eqnarray}where the integral denotes
\begin{equation}
\int_{\mathbf{\hat{k}}_{i\ell}} = \int_{\mathbf{k}\in k_i}\frac{\mathbf{dk}}{V_{k_i}}(2\ell+1){\cal L}_\ell(\mu_\mathbf{k}).
\end{equation}
The tree-level covariance after reconstruction can be obtained by replacing $Z_n$ with the kernel of post-reconstructed (post-rec) spectra $Z_n^{\rm (rec)}$ derived by \citep{HKT19}. The first-order kernel is not changed by the reconstruction
\begin{equation}
Z_1^{\rm (rec)}(\mathbf{k})=Z_1(\mathbf{k}).
\end{equation}
The relation of the 2nd and 3rd-order post-rec kernels to the pre-reconstructed (pre-rec) kernels are given as \citep{HKT19}

\begin{eqnarray}
Z_2^{\rm (rec)}(&\mathbf{k}_1&,\mathbf{k}_2)=Z_2(\mathbf{k}_1,\mathbf{k}_2) \nonumber \\
&&+\frac12\left[(\mathbf{k}\cdot \mathbf{S}^{\rm z(1)}(\mathbf{k}_1))
(\mathbf{k}_2\cdot \mathbf{L}^{z(1)}(\mathbf{k}_2))\right. \nonumber \\
&& +\left.(\mathbf{k}\cdot \mathbf{S}^{\rm z(1)}(\mathbf{k}_2))
(\mathbf{k}_1\cdot \mathbf{L}^{z(1)}(\mathbf{k}_1))\right],
\label{eq:Z2_recon}
\end{eqnarray}
and
\begin{eqnarray}
Z_3^{\rm (rec)}(&\mathbf{k}_1&,\mathbf{k}_2,\mathbf{k}_3)=
Z_3(\mathbf{k}_1,\mathbf{k}_2,\mathbf{k}_3) \nonumber \\
&&+\frac16\left[2(\mathbf{k\cdot S}^{z(1)}(\mathbf{k}_1))Z_2(\mathbf{k}_2,\mathbf{k}_3)\right.
\nonumber \\
&&+(\mathbf{k\cdot S}^{z(1)}(\mathbf{k}_1))(\mathbf{k\cdot S}^{z(1)}(\mathbf{k}_2))(\mathbf{k_3}\cdot\mathbf{L}^{z(1)}(\mathbf{k_3}))
\nonumber \\
&& +(\mathbf{k\cdot S}^{z(2)}(\mathbf{k}_1,\mathbf{k}_2))
(\mathbf{k_3}\cdot\mathbf{L}^{z(1)}(\mathbf{k_3}))
\nonumber \\
&&+\left.{\rm (2~perms.)}
\right],
\label{eq:Z3_recon}
\end{eqnarray}
where $\mathbf{k}=\mathbf{k}_1+...+\mathbf{k}_n$ in the $n$-th order kernel.
In the above equations, $\mathbf{L}^{z(n)}$ represents the $n$-th order Lagrangian kernel in redshift space, which is related to the same order of Lagrangian kernel in real space $\mathbf{L}^{(n)}$  as \citep{Matsubara08a}
\begin{equation}
\mathbf{L}^{z(n)}=\mathbf{R}^{(n)}\mathbf{L}^{(n)}.
\end{equation}
The redshift-space distortion tensor at $n$-th order $\mathbf{R}^{(n)}$ is given by
\begin{equation}
\mathbf{R}^{(n)}_{ij}=\delta_{ij}+nf\hat{z}_i\hat{z}_j,
\end{equation}
where $\delta_{ij}$ is the Kronecker delta and $\hat{z}_i$ the i-th component of the unit vector in the line-of-sight direction. In a standard reconstruction \citep{Eisenstein07b}, the shift field is estimated from the smoothed density field using the inverse Zeldovich approximation. The $n$-th order kernel of the shift field $\mathbf{S}^{z(n)}$ is then given by
\begin{equation}
\label{eq:shiftkernel}
\mathbf{S}^{z (n)}(\mathbf{k}_1,...,\mathbf{k}_n)=
-n!W(k)\mathbf{L}^{(1)}(\mathbf{k})Z_n(\mathbf{k}_1,...,\mathbf{k}_n),
\end{equation}
where $W(k)$ is the smoothing kernel and we adopt Gaussian kernel, i.e., $W(k)=\exp(-k^2R_s^2/2)$ with different smoothing scales $R_s=5h^{-1}$Mpc, $10h^{-1}$Mpc and 20$h^{-1}$Mpc.

Figure \ref{fig:cov_tree} compares the tree-level non-Gaussian covariance of the monopole spectra before and after the reconstruction. The plotted covariance is normalized with their Gaussian components, i.e., ${\bf Cov}_{00}^{\rm (tree)}(k_i,k_j)/[{\bf Cov}_{00}^{\rm (G)}
(k_i,k_i){\bf Cov}_{00}^{\rm (G)}(k_j,k_j)]^{1/2}$ . The off-diagonal components have positive values for the pre-rec spectra, which means that the different modes are positively correlated by the mode coupling of gravity. After reconstruction, we find that the positive correlation decreases and becomes negative at $R_s$ less than 10$h^{-1}$Mpc. This comes from that the values of the tree-level trispectra shift from positive to negative by replacing the perturbative kernels with the reconstructed one. This is related to our previous finding that the amplitudes of the one-loop terms of the power spectrum given by the same perturbative kernels decrease after the reconstruction \citep{HKT19}. 
\begin{figure*}
\begin{center}
\includegraphics[width=16cm]{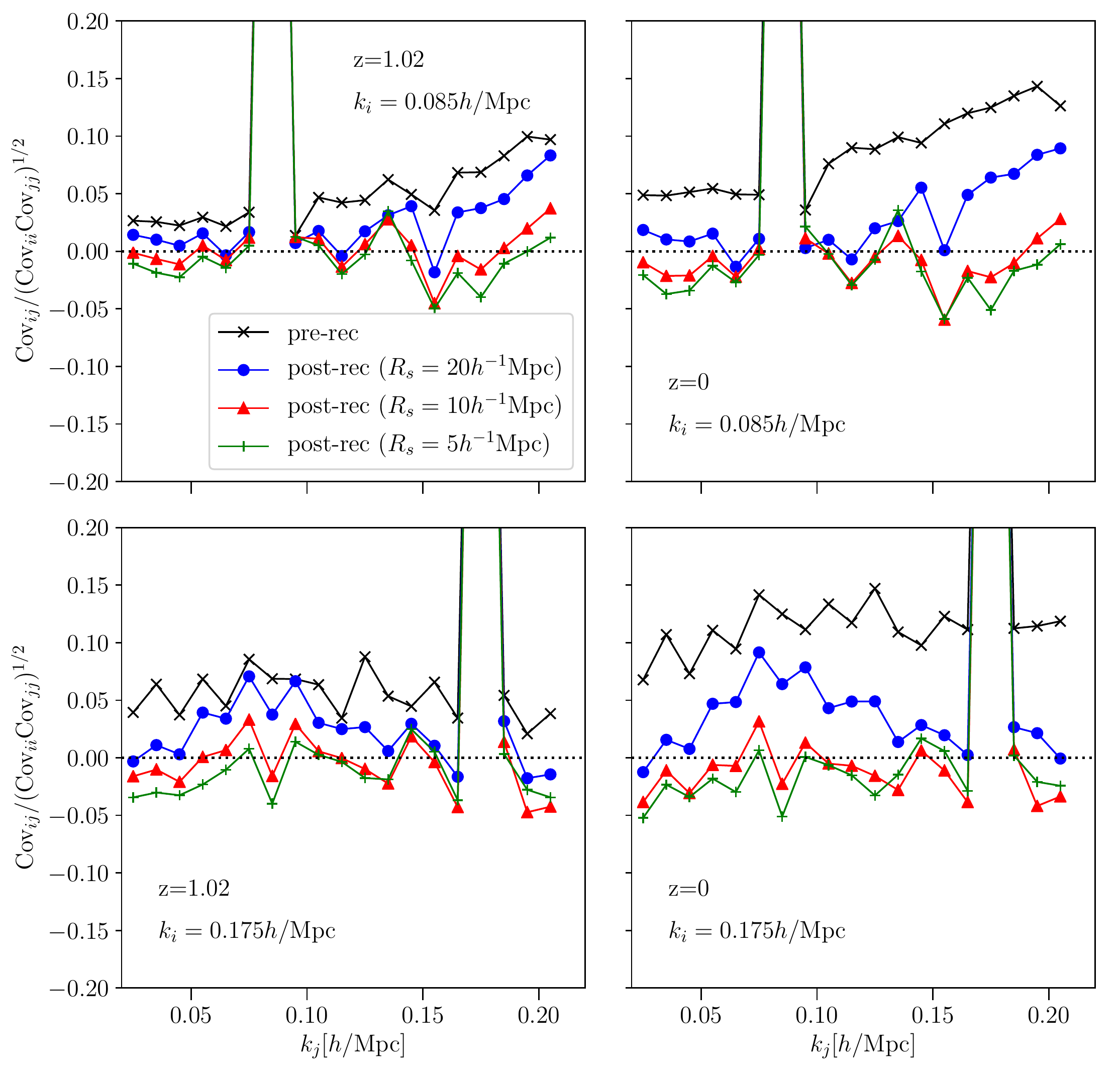}
\caption{Correlation coefficients of the monopole components of the matter power spectrum with fixed $k_i=0.085h$/Mpc (upper) and 0.175$h$/Mpc (lower) at $z=1.02$ (left) and $z=0$ (right). Different symbols denote the results for pre-rec and post-rec spectra with different $R_s$. Off-diagonal components of post-rec spectra significantly decreases and have negative values at $R_s$ less than $10h^{-1}$Mpc, which is consistent with the behavior of the perturbation theory in Figure \ref{fig:cov_tree}.}
\label{fig:cov_offdiag}
\end{center}
\end{figure*}
\section{N-body simulations}
\label{sec:nbody}

We measure the covariance of the multipole components of matter power spectra over an ensemble of dark matter $N$-body simulations as follows:
\begin{eqnarray}
\label{eq:covsim}
{\bf Cov}_{\ell\ell^\prime}(&k&,k^\prime)=\frac{1}{N_{\rm real}-1}
~~~~~~~~~~~~~~~~~~~~~~~~~~~~~~~~~~~~
\nonumber \\
&&\times\sum_i^{N_{\rm real}}
[P_{\ell,i}(k)-\bar{P}_\ell(k)][P_{\ell^\prime,i}(k^\prime)-\bar{P}_{\ell^\prime}(k^\prime)],
\end{eqnarray}
where $N_{\rm real}$  is the number of realizations and $\bar{P}_\ell$ is the averaged multipole components of matter power spectra. The binning width of $k$ is uniformly set to be $0.01h$/Mpc and the minimum value of $k$ is set to be 0.01$h$/Mpc. We perform N-body simulations using a publicly available code {\tt Gadget-2} \citep{Springel05}. The initial distribution of mass particles is based on the 2LPT code \citep{CPS2006,Nishimichi2009} with Gaussian initial conditions at the input redshift of $31$. The initial redshift does not affect (less than $1 \%$) the power spectrum at $k<0.3h$/Mpc and $z=0$ (see Fig. 35 of \cite{Nishimichi2019}). The initial linear power spectrum is computed by CAMB \citep{Lewis2000}. We adopt 4000 realizations with 512$^3$ mass particles in a cubic box with a side length of $500h^{-1}$Mpc and two output redshifts of $z=0$ and $z=1.02$.

The $N$-body particles are assigned to $512^3$ grid cells with the clouds-in-cell (CIC) method to calculate the density contrast. We then perform the Fourier transform \footnote{FFTW3 at http://www.fftw.org} to measure the multipole components of the power spectra $P_\ell$ after the pixel window effect is corrected \citep[e.g.,][]{Jing05}. The reconstructed density field is computed as follows \citep{Eisenstein07b}:
\begin{itemize}
\item The shift field for the reconstruction is computed from the smoothed redshift-space mass density field using the inverse Zeldovich approximation, i.e., $\tilde{\mathbf{s}}(\mathbf{k})=-(\mathbf{k}/k^2)\tilde\delta_m^{\rm (z)}(\mathbf{k}) W(k)$ with Gaussian smoothing kernel $W(k)=\exp{(-k^2R_s^2/2)}$ at  $R_s$=5$h^{-1}$Mpc, 10$h^{-1}$Mpc and $20h^{-1}$Mpc. Note that we leave the reconstructed field anisotropic on large scales to constrain the growth rate from the anisotropy due to the redshift-space distortion.
\item Each mass particle is displaced following the above shift field at the position interpolated from the shift field at neighboring grids with the CIC scheme.
\item Random particles are also displaced using the same shift vector field in the same manner as the mass particles.
\item Reconstructed density field is obtained by the displaced random field subtracted from the displaced data field as $\delta^{\rm (rec)}=\delta_{\rm d}^{\rm (rec)}-\delta_{\rm r}^{\rm (rec)}$
\end{itemize} 

We also compute the matter power spectra from 8 realizations of large-box $N$-body simulations with a side length of 4$h^{-1}$Gpc. Each realization contains $4096^3$ mass particles and they are assigned to 2048$^3$ grid cells with the CIC method. We confirm that the averaged spectra from $500h^{-1}$Mpc box is consistent with those from $4h^{-1}$Gpc box, however, we find a large fluctuation of the quadrupole spectrum from $500h^{-1}$Mpc box due to a low resolution in $k$-space. We therefore adopt the power spectrum from 4$h^{-1}$Gpc-box simulations and add the following correction to the covariance from $500h^{-1}$Mpc box simulations as
\begin{eqnarray}
\label{eq:cov_sim2}
{\bf Cov}_{\ell\ell^\prime}(&k&,k^\prime)= {\bf Cov}_{\ell\ell^\prime}^{\rm (500h^{-1}Mpc)}(k,k^\prime)~~~~~~~~~~~~~~~~~~~~~~~~
\nonumber \\ 
&&\times
\left[\frac{\bar{P}_\ell^{\rm (4h^{-1}Gpc)} (k)}{\bar{P}_\ell^{\rm (500h^{-1}Mpc)}(k)}\right]\left[\frac{\bar{P}_{\ell^\prime}^{\rm (4h^{-1}Gpc)} (k^\prime)}{\bar{P}_{\ell^\prime}^{\rm (500h^{-1}Mpc)}(k^\prime)}\right],
\end{eqnarray}
where $\bar{P}_\ell^{(L_{\rm box})}$ is the averaged multipole power spectrum from $N$-body simulations with a side length of $L_{\rm box}$.

Figure \ref{fig:cov_offdiag} shows the correlation matrix of the monopole spectra computed from the simulations at fixed $k_i=0.085h$/Mpc and 0.175$h$/Mpc at $z=1.02$ and $0$. Pre-rec spectra are positively correlated among different modes and thus the off-diagonal components are positive \citep{Scoccimarro99b,Meiksin99}. We find that the off-diagonal components substantially decrease to be nearly zero by reconstruction with $R_s=10h^{-1}$Mpc. At $R_s=5h^{-1}$Mpc, the off-diagonal components become negative values. This behavior is qualitatively consistent with the perturbation theory shown in the previous section.

\begin{figure*}
\begin{center}
\includegraphics[width=8.5cm]{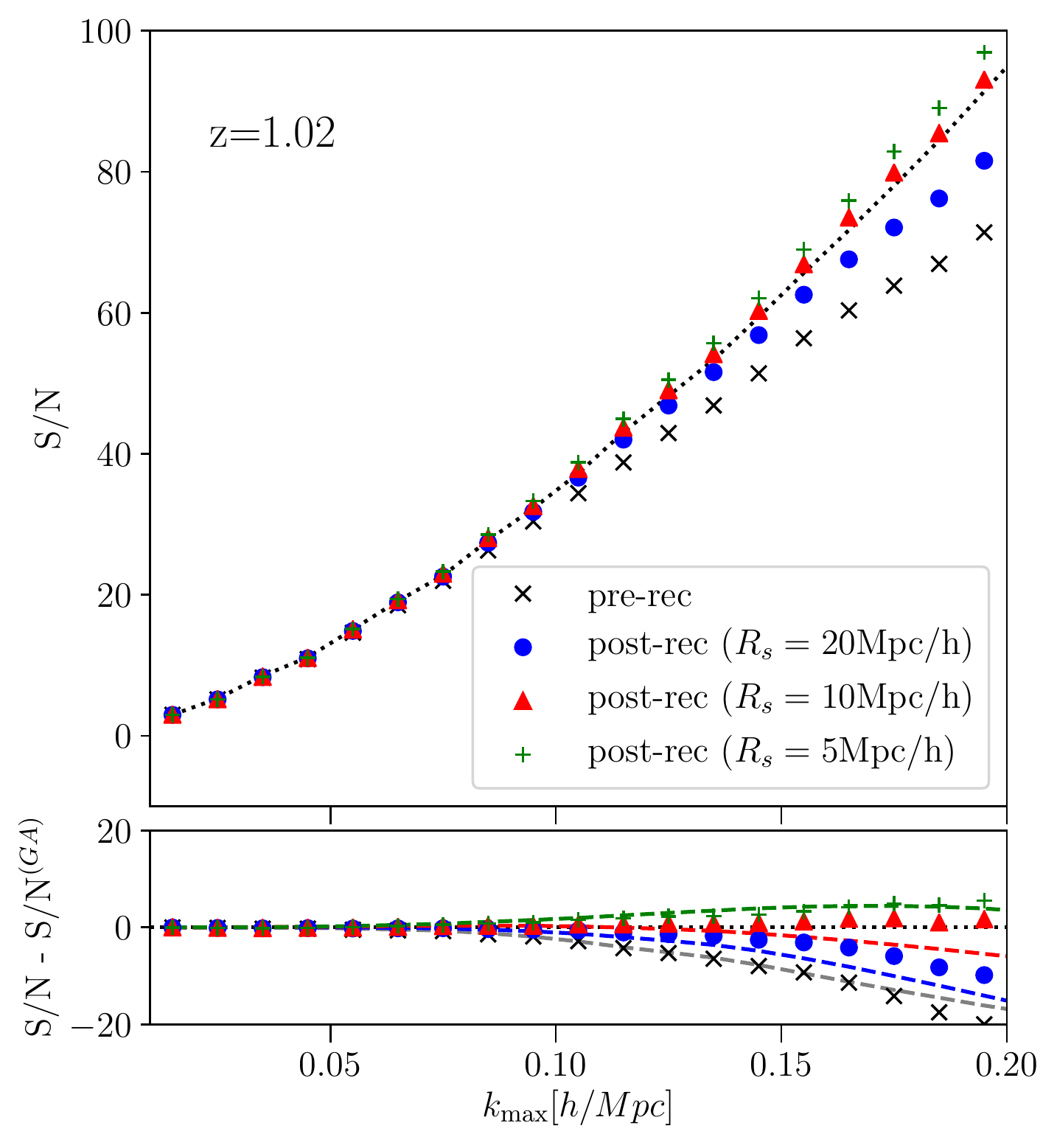}
\includegraphics[width=8.5cm]{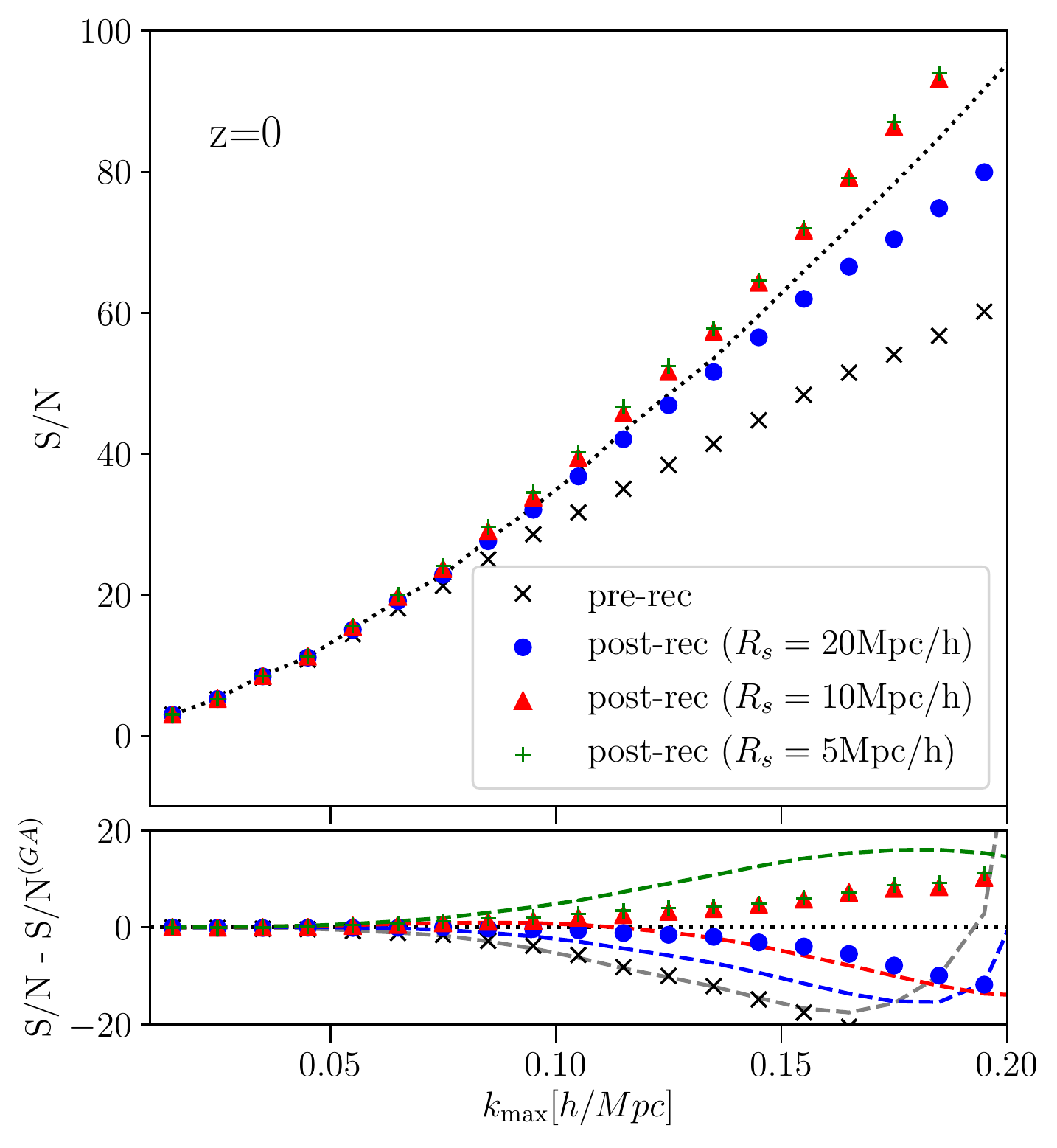}
\caption{Comparison of the signal-to-noise ratios (S/N) of the sum of the monopole and quadrupole components of redshift-space matter power spectra before reconstruction and after reconstruction with $R_s=20h^{-1}$Mpc, $10h^{-1}$Mpc and 5$h^{-1}$Mpc  at $z=1.02$ (left) and $z=0$ (right). For comparison, we plot the S/N of the linear power spectrum using the analytical Gaussian covariance (Cov$^{\rm (GA)}$) with dotted lines. We find that the post-rec spectra has a better S/N than pre-rec one and also that the post-rec spectra with $R_s=10h^{-1}$Mpc and 5$h^{-1}$Mpc have higher S/N than the Gaussian one.  Lower panels focus on the differences of the S/N for the linear Gaussian one. For comparison, the S/N of the linear power spectra using the tree-level perturbative covariance (Cov$^{\rm (GA)}$+Cov$^{\rm (tree)}$) are plotted with dashed lines and show the similar behavior.}
\label{fig:pkSN}
\end{center}
\end{figure*}

\begin{figure*}
\begin{center}
\includegraphics[width=8.5cm]{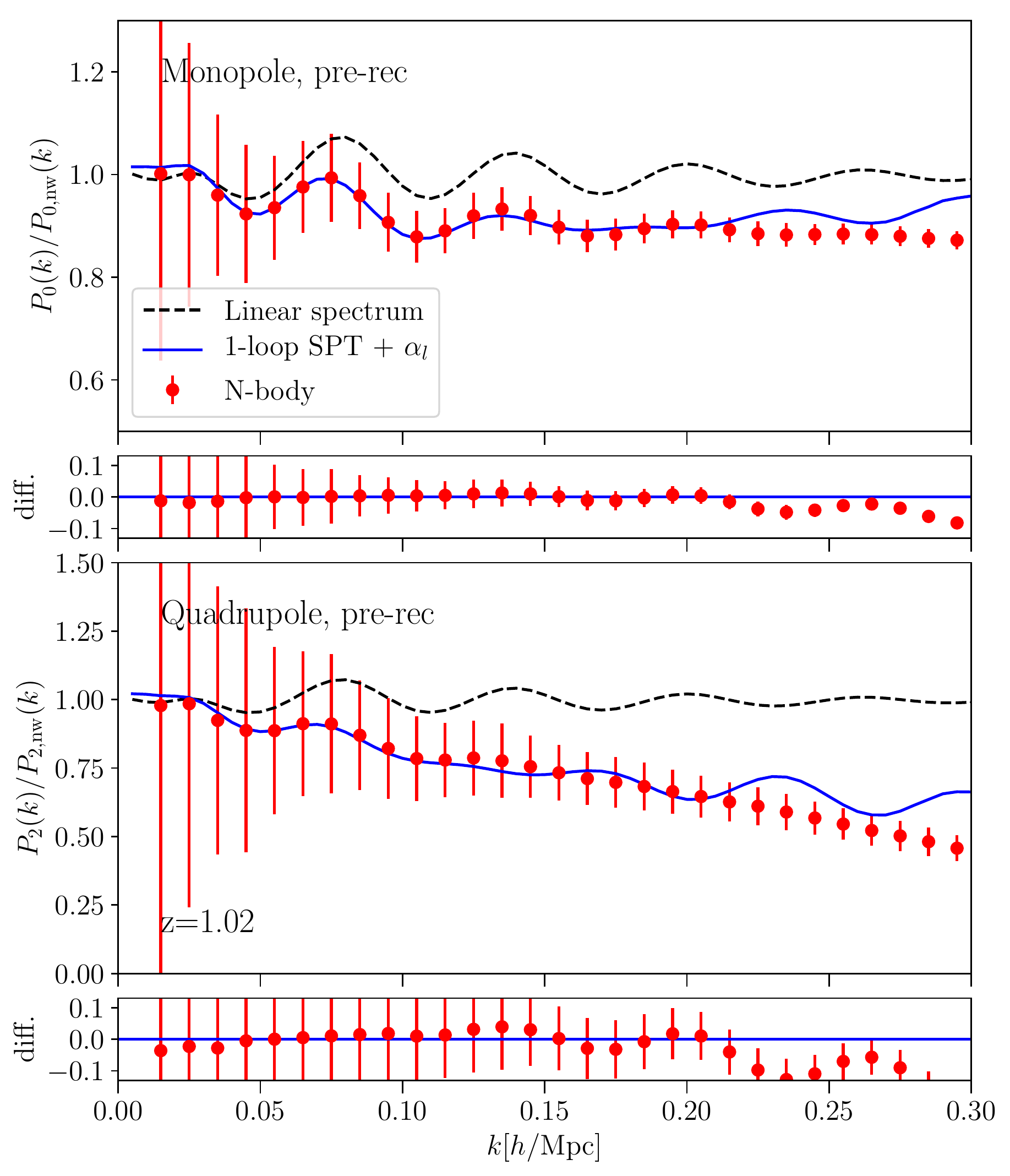}
\includegraphics[width=8.5cm]{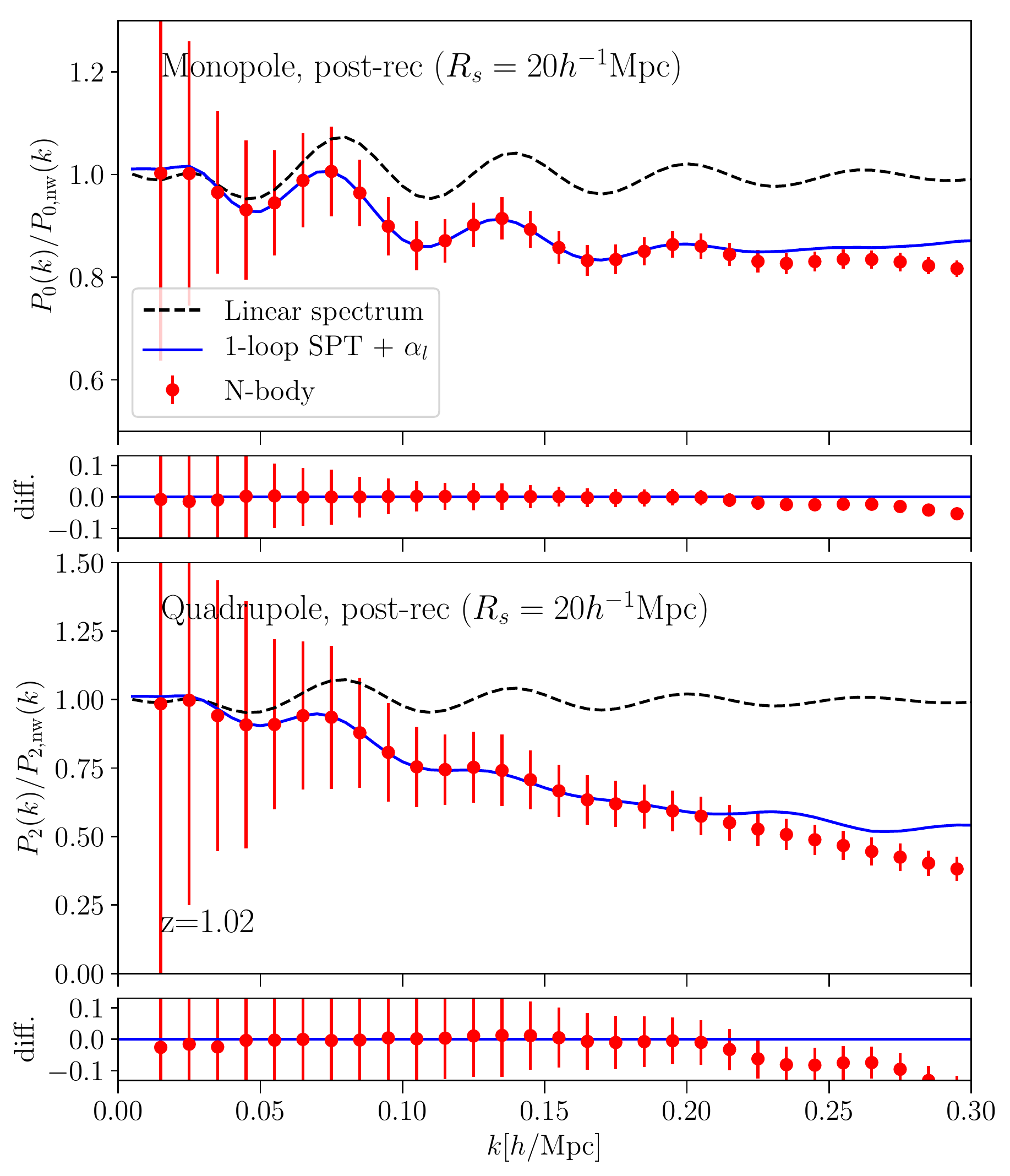}
\includegraphics[width=8.5cm]{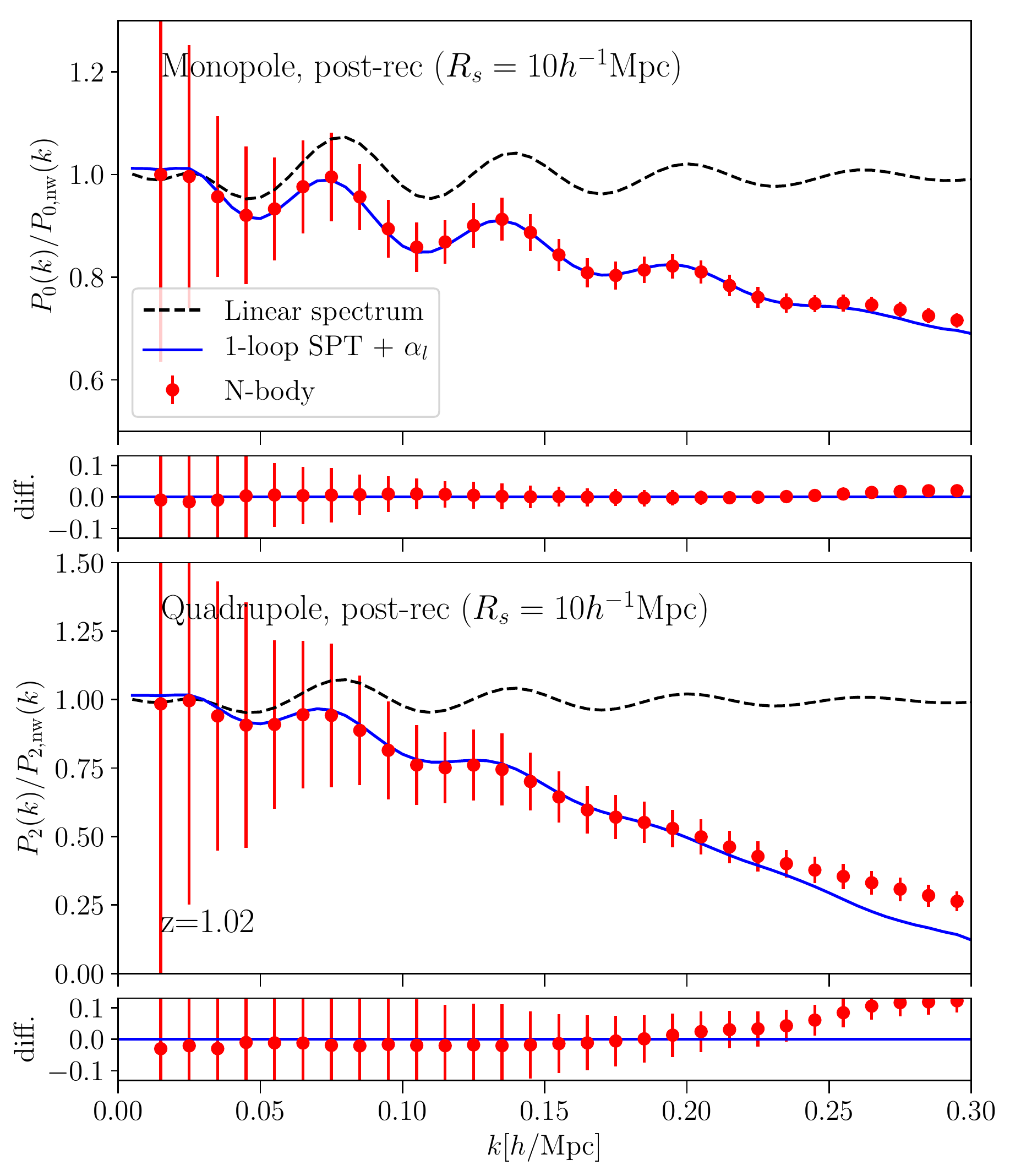}
\includegraphics[width=8.5cm]{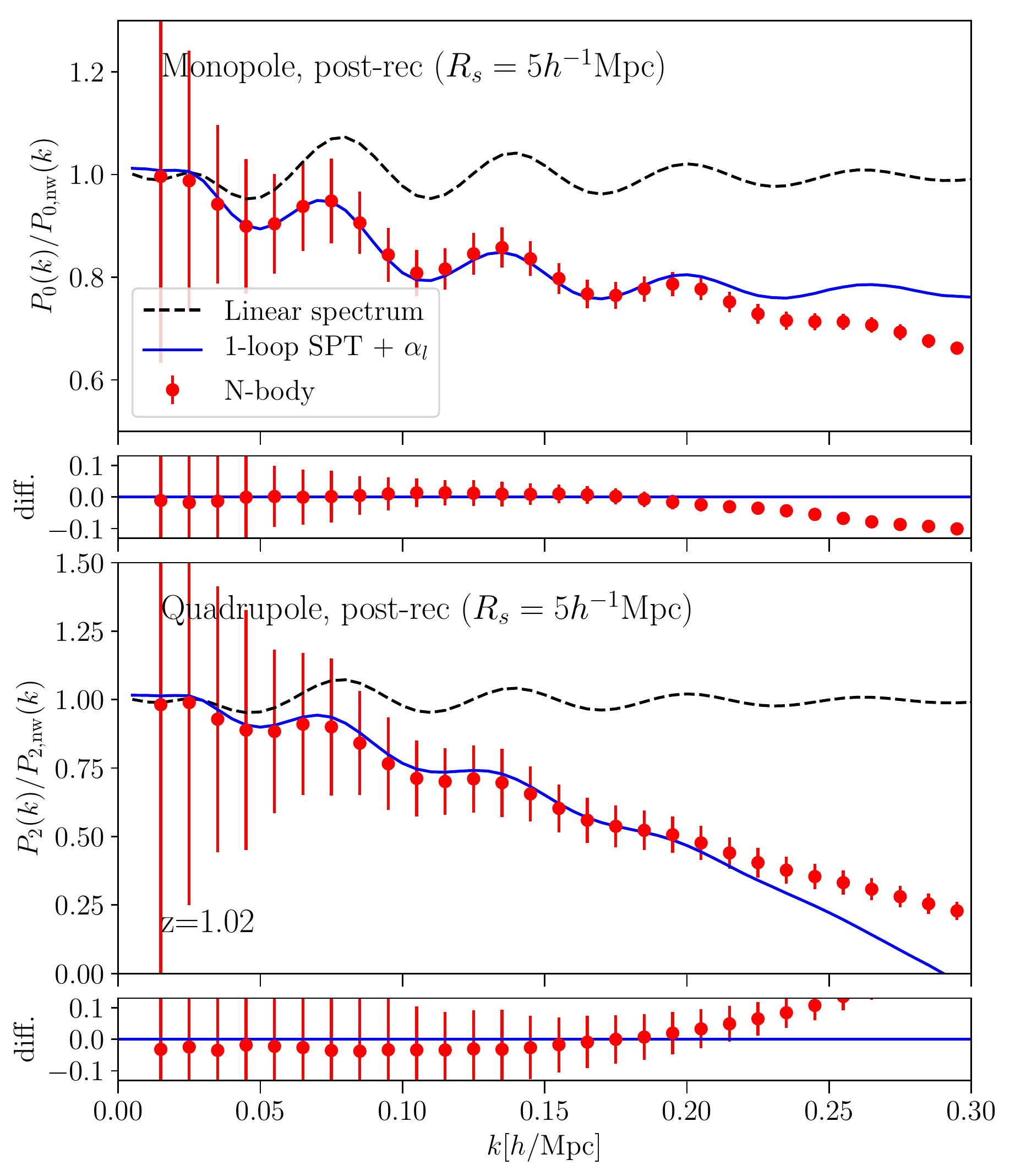}
\caption{Monopole and quadruple components of the matter power spectrum from N-body simulations (filled circles) before reconstruction (upper-left) and after reconstruction with $R_s=20h^{-1}$Mpc (upper-right), 10$h^{-1}$Mpc (lower-left) and 5$h^{-1}$Mpc (lower-right). Error-bars denote the 1$\sigma$ sample variance from 4000 $N$-body simulations where each volume is ($500h^{-1}$Mpc)$^3$. For comparison, the 1-loop perturbative formula are plotted with solid lines using the best-fit values of the lowest-order counterterms $\alpha_\ell$  ($\ell=0,2$) up to $k_{\rm max}=0.2h$/Mpc. The linear power spectra are also plotted with dashed lines. All of the plotted spectra are divided with the no-wiggle spectra. Small panels show the differences between the simulated spectra and the 1-loop perturbative formula with the best-fit values of $\alpha_\ell$. The output redshift is $z=1.02$.}
\label{fig:pkl_fit_z1}
\end{center}
\end{figure*}

\begin{figure*}
\begin{center}
\includegraphics[width=8.5cm]{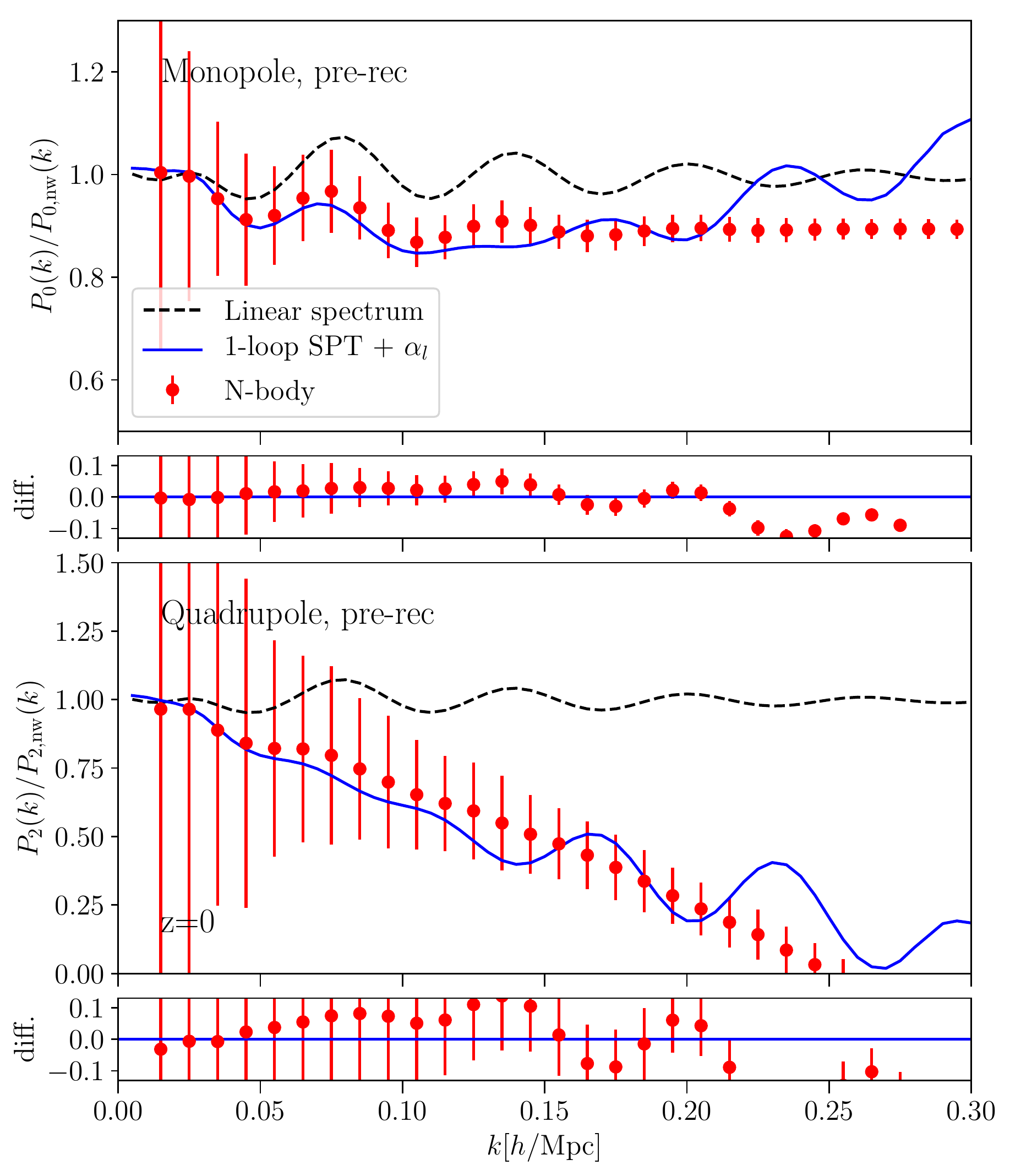}
\includegraphics[width=8.5cm]{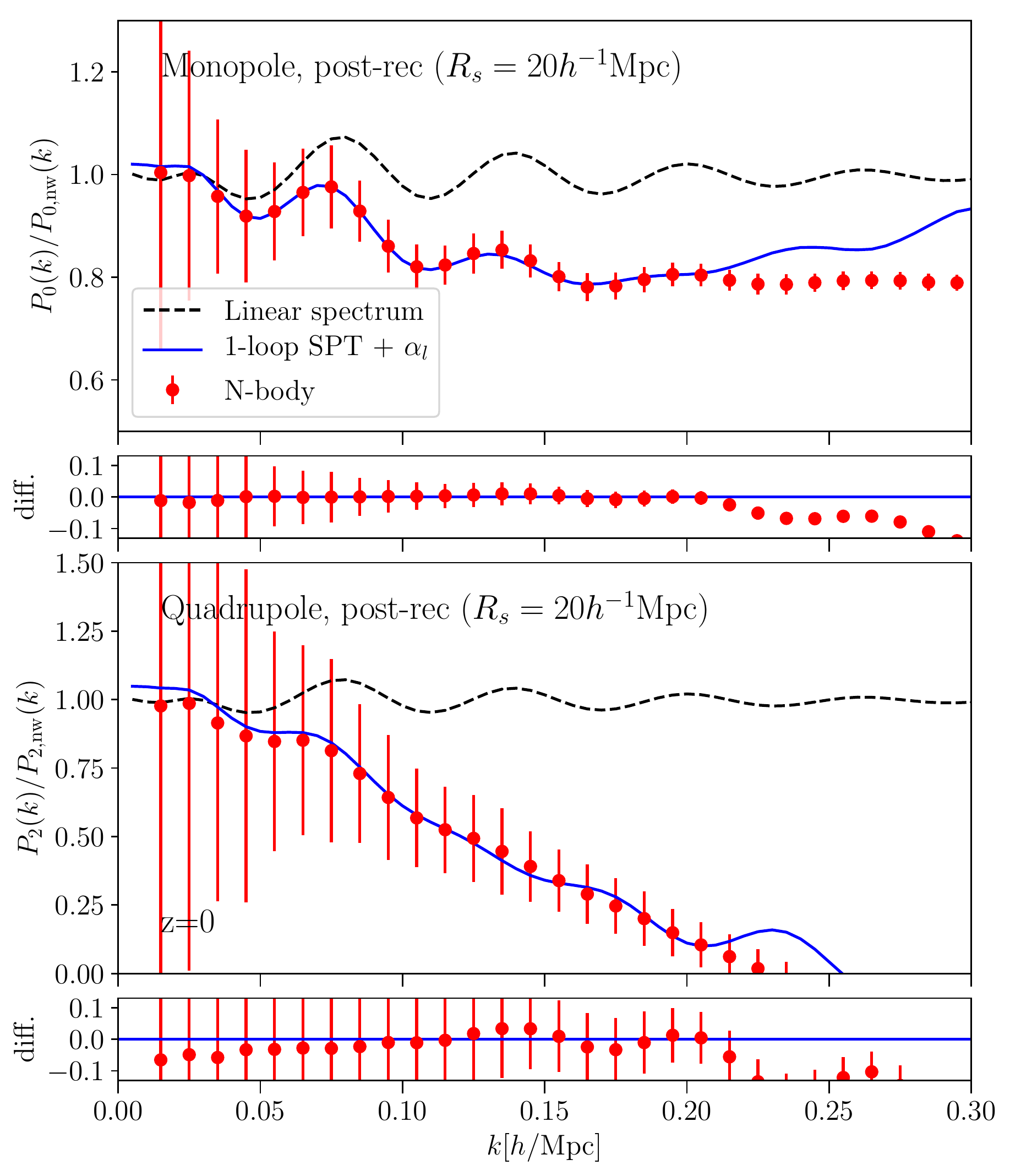}
\includegraphics[width=8.5cm]{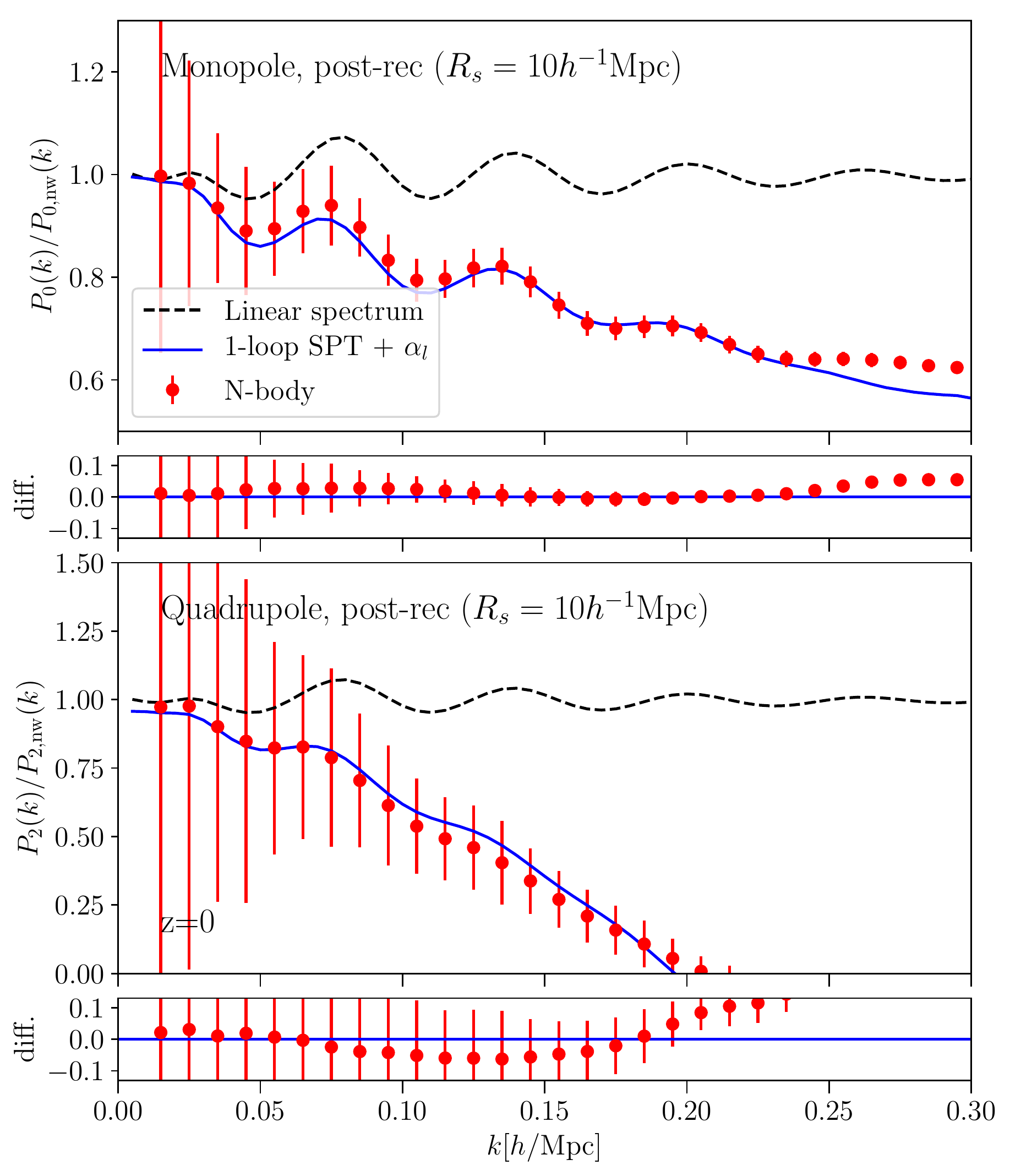}
\includegraphics[width=8.5cm]{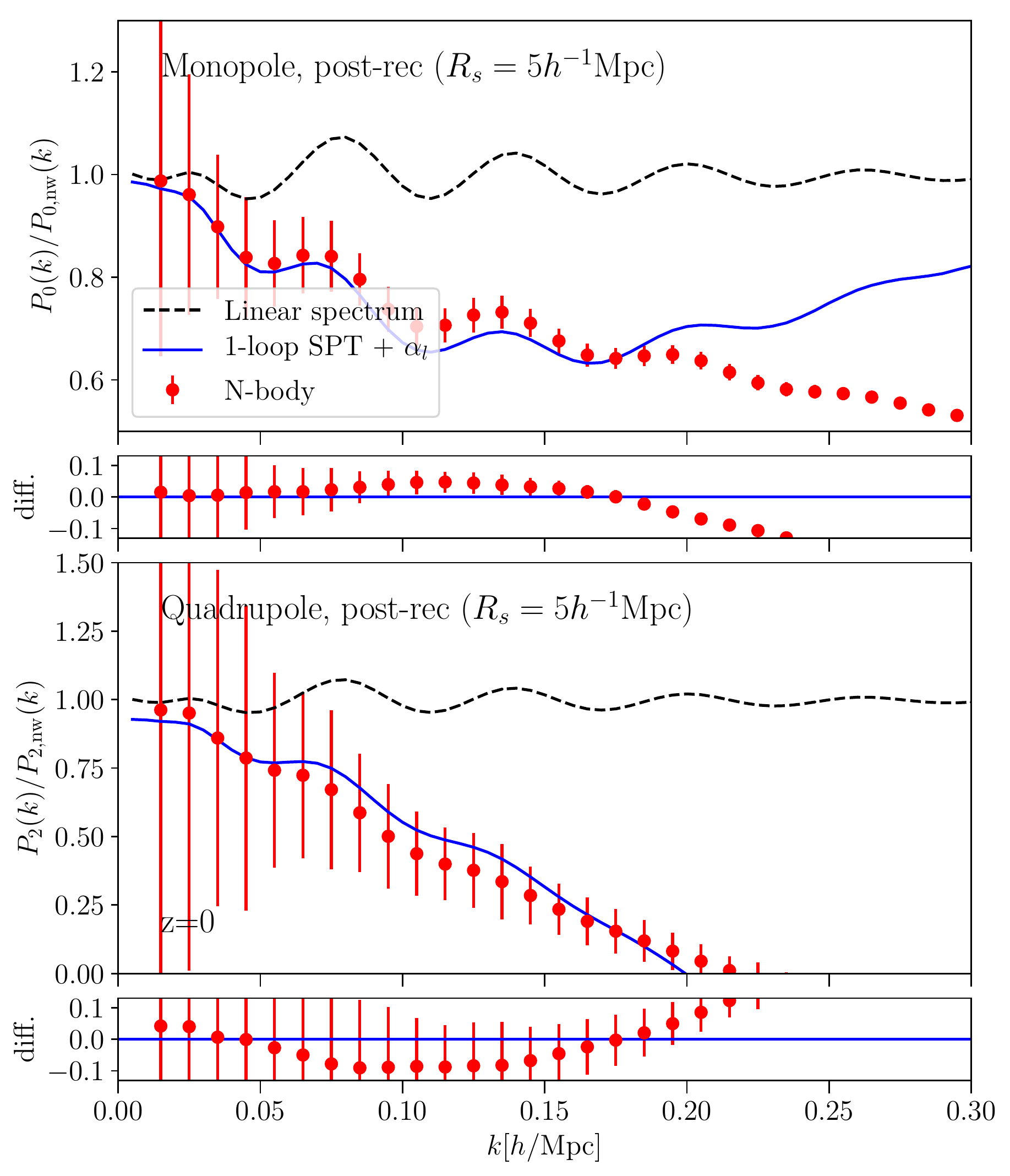}
\caption{Same as Figure \ref{fig:pkl_fit_z1} but for the output redshift of $z=0$.}
\label{fig:pkl_fit_z0}
\end{center}
\end{figure*}

\begin{figure*}
\begin{center}
\includegraphics[width=17cm]{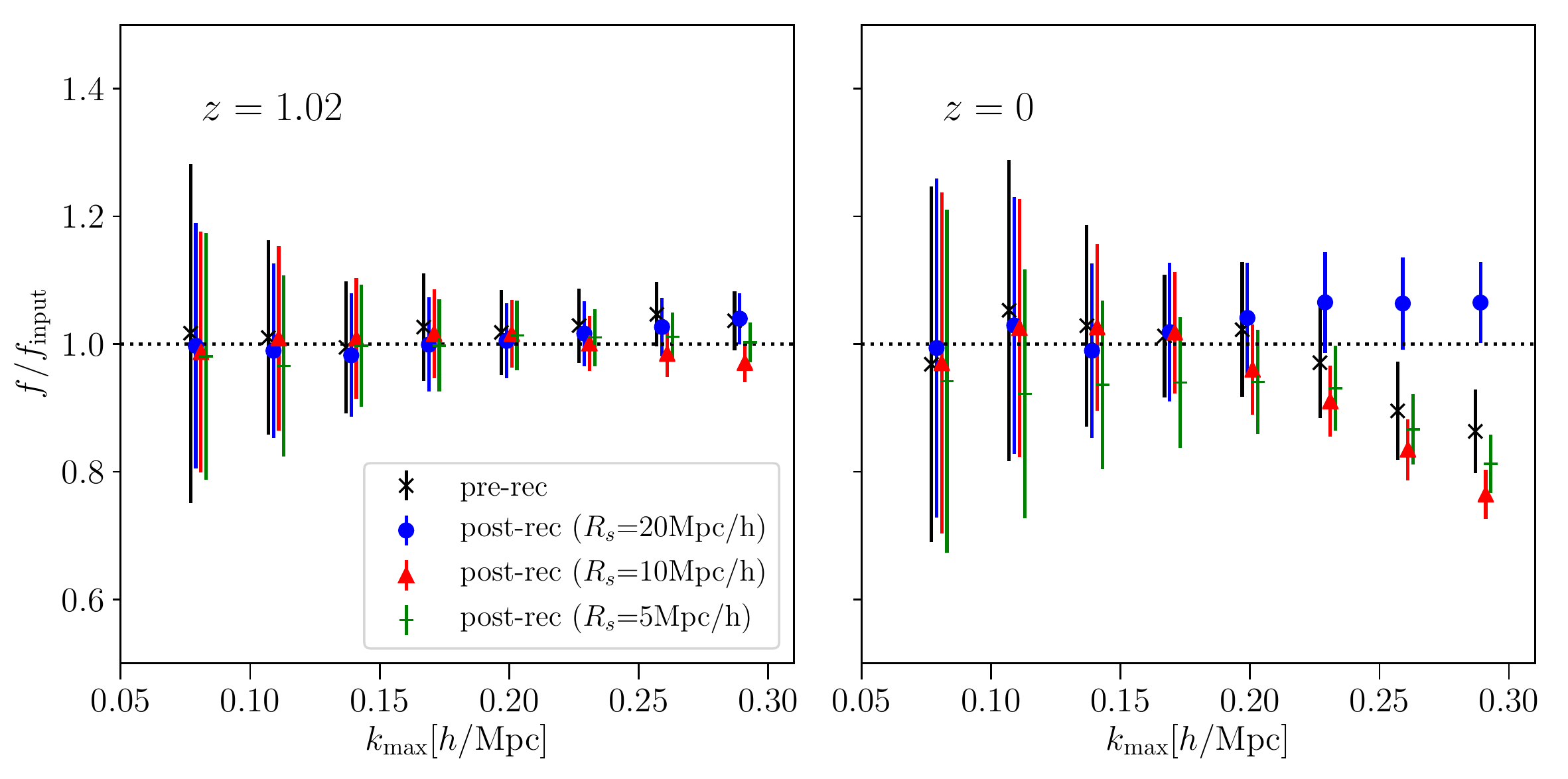}
\caption{Linear growth rate relative to the input value obtained by fitting the 1-loop perturbative formulae of the monopole and quadrupole power spectra to the simulated spectra with the maximum wavenumber $k_{\rm max}$ varied (eq.[\ref{eq:likelihood}]). The counterterms $\alpha_\ell$ ($\ell=0$ and 2) are freely fitted, while other cosmological parameters are fixed. Different plots show the pre-rec spectra (x-shaped crosses) and post-rec spectra with $R_s=20h^{-1}$Mpc (circles), 10$h^{-1}$Mpc (triangles) and 5$h^{-1}$Mpc (+-shaped crosses) at $z=1.02$ (left panels) and $z=0$ (right panels). Covariance of the multipole power spectra is estimated from 4000 realizations of $N$-body results with $(500h^{-1}{\rm Mpc})^3$ volume. The error-bars denote the $1\sigma$ uncertainty.}
\label{fig:fz_err}
\end{center}
\end{figure*}

\section{Signal-to-noise ratio of multipole power spectra}
\label{sec:SN}
In this section, we evaluate the information content of redshift-space matter power spectra from the following signal-to-noise ratio (S/N):
\begin{equation}
\label{eq:sn}
{\rm (S/N)^2}=\sum_{\ell,\ell^\prime}^{0,2}\sum_{i,j}^{k_i,k_j\le k_{\rm max}}P_\ell(k_i)({\bf Cov}^{-1})_{\ell\ell^\prime}(k_i,k_j)P_{\ell^\prime}(k_j).
\end{equation}
The multipole spectra $P_\ell$ and their covariance are directly estimated from the simulations as explained in the previous section. The inverse covariance matrix is computed by multiplying a so-called Hartlap factor $\alpha=(N_{\rm real}-N_{\rm bin}-2)/(N_{\rm real}-1)$ with the inverse of the covariance matrix (eq.
[\ref{eq:covsim}])\citep{Hartlap07}.

Figure \ref{fig:pkSN} compares the S/N of pre-rec spectra and post-rec spectra with different $R_s$ as a function of the maximum wavenumber $k_{\rm max}$. We find that the post-rec spectra have higher S/N than the pre-rec spectra. The improvement is larger at higher $k$. For example, the S/N of the post-rec spectra with $R_s=10h^{-1}$Mpc is improved by 7\% ($k_{\rm max}=0.1h$/Mpc) and 30\% ($k_{\rm max}=0.2h$/Mpc) at $z=1.02$ relative to the pre-rec spectra. The improvement is more significant at $z=0$: 18\% ($k_{\rm max}=0.1h$/Mpc) and 69\% ($k_{\rm max}=0.2h$/Mpc). Since the diagonal components of the covariance matrix is dominated by the Gaussian terms, the improvement of the S/N mainly comes from the decrement of the off-diagonal components as shown in Figure \ref{fig:cov_tree} and \ref{fig:cov_offdiag}. 

We also plot the S/N estimated from the linear spectra and the Gaussian covariance as a reference. Lower panels focus on the differences of S/N from the linear Gaussian one. Interestingly, it is found that the S/N of the post-rec spectra at $R_s=10h^{-1}$Mpc and 5$h^{-1}$Mpc are comparable to or higher than the linear Gaussian one. The similar trend is found from the perturbative predictions where the linear spectra and the tree-level covariance (eq.\ref{eq:cov_tree}) are applied to calculate the S/N (eq.\ref{eq:sn}), though the agreement of the perturbation with the numerical results is limited to be at $k\le 0.1h$/Mpc. The S/N from the perturbation rapidly increase at high $k$ because the determinant of tree-level covariance diverges \citep{Takahashi09}. As shown in Figure \ref{fig:cov_tree} and \ref{fig:cov_offdiag}, the off-diagonal components become negative at $R_s\simlt 10 h^{-1}$Mpc and thereby the S/N of the post-rec spectra becomes higher than the linear Gaussian one.

Information of nonlinear growth of structure can be normally captured by higher-order statistics beyond two-point statistics. The reconstruction returns the information leaking to higher-order statistics back to the two-point statistics. The return is larger at smaller $R_s$ where smaller scales can be reconstructed and thereby the S/N increases at smaller $R_s$. The growth information is however buried on strongly nonlinear regime and thus the increment of S/N from $R_s=10h^{-1}$Mpc to $R_s=5h^{-1}$Mpc at $z=0$ is limited.

\begin{figure*}
\begin{center}
\includegraphics[width=8.5cm]{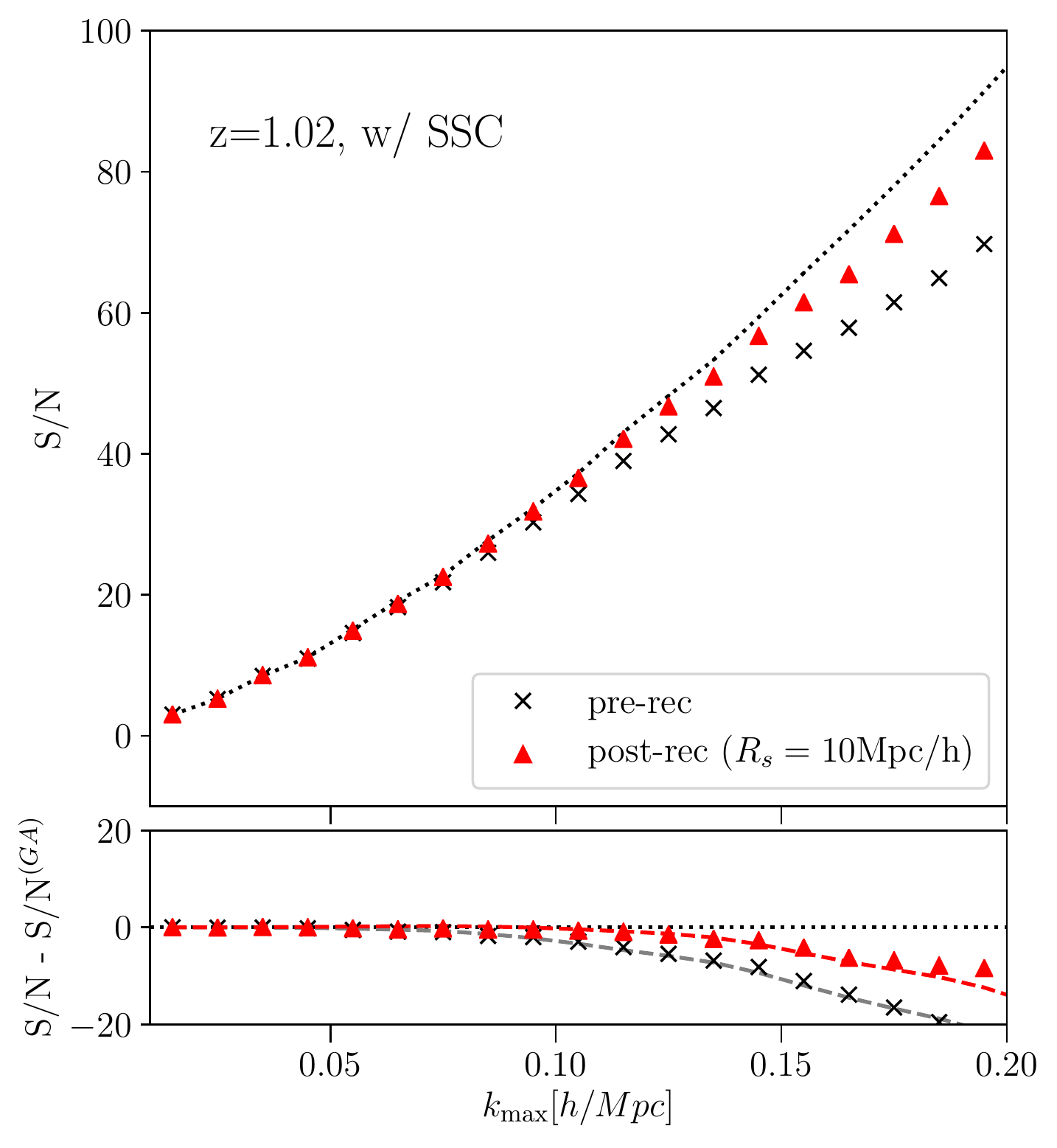}
\includegraphics[width=8.5cm]{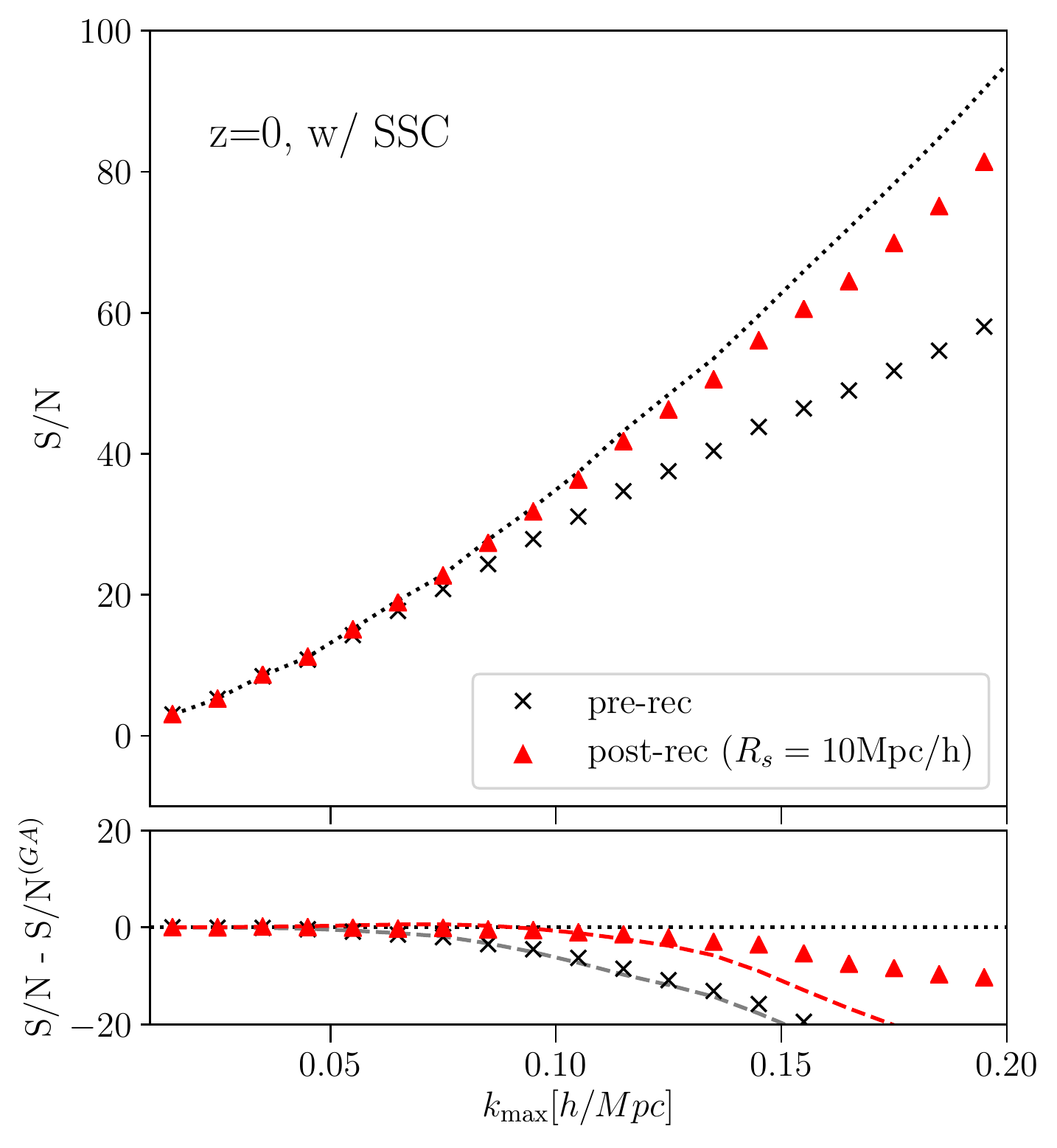}
\caption{Same as Figure \ref{fig:pkSN} but for the S/N of pre-rec and post-rec spectra with $R_s=10h^{-1}$Mpc and the super-sample covariance (SSC) is included in the simulated covariance. The predictions from the 1-loop perturbative formulae are also estimated from the covariance including SSC, i.e.,  Cov$^{\rm (GA)}$+Cov$^{\rm (tree)}$+Cov$^{\rm (SSC)}$ . }
\label{fig:pkSN_ssc}
\end{center}
\end{figure*}

\begin{figure*}
\begin{center}
\includegraphics[width=17cm]{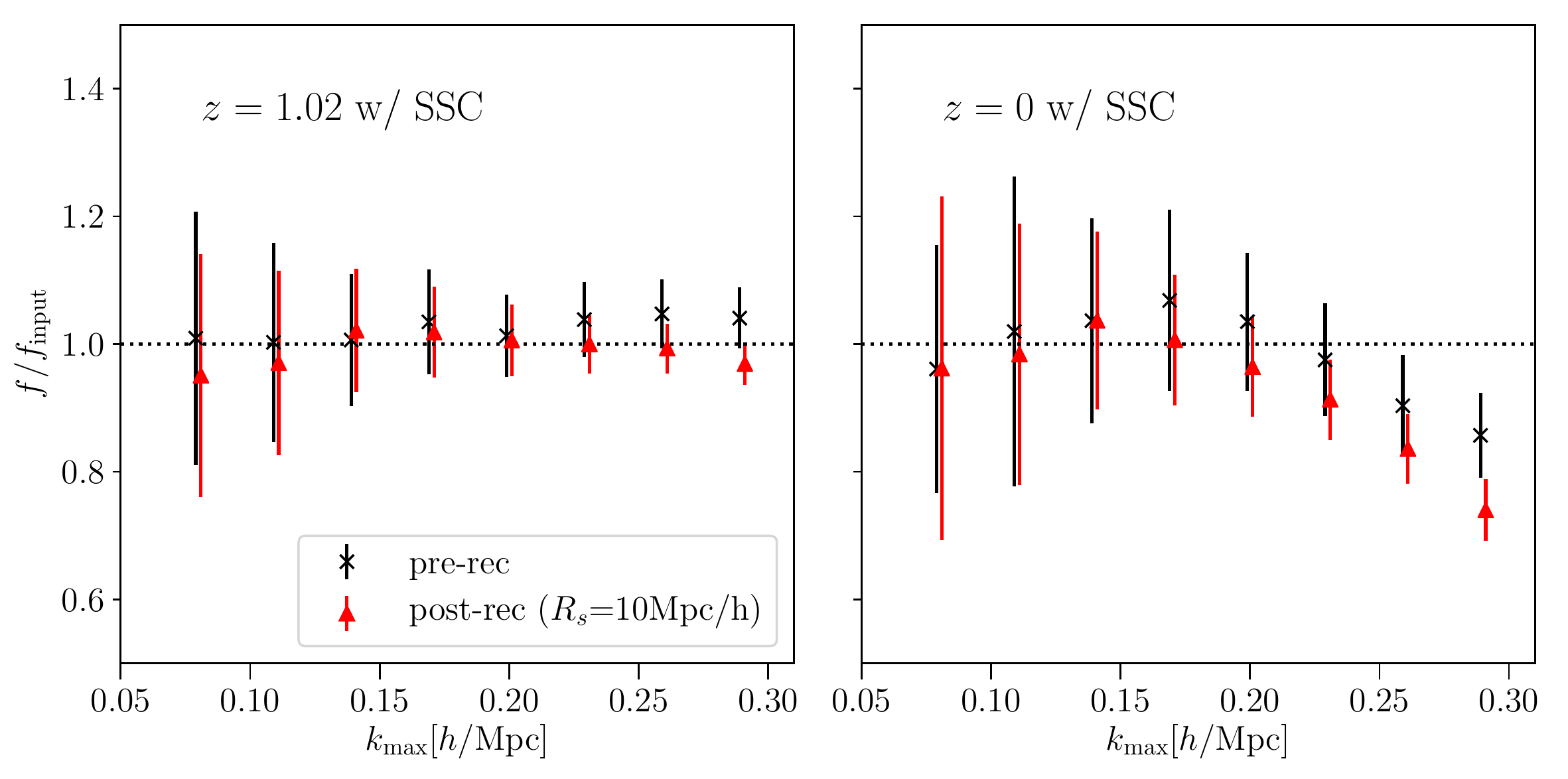}
\caption{Same as Figure \ref{fig:fz_err} but for the SSC effect is included. Here we fix the smoothing scale $R_s$ for the post-rec spectra 10$h^{-1}$Mpc. }
\label{fig:fz_err_ssc}
\end{center}
\end{figure*}

\section{Impacts on growth rate measurements}
\label{sec:growthrate}
In this section we explore if the estimates of cosmological parameters are improved by reconstruction, particularly focusing on the growth rate. We evaluate the error of the growth rate including the systematics when using the 1-loop perturbative formulae as a theoretical modeling of the matter power spectra. The likelihood is estimated as follows:
\begin{eqnarray}
\label{eq:likelihood}
{\cal L}&\propto &\exp{\left(-\frac{\chi^2}{2}\right)}, \\
\chi^2 (\mathbf{p})&=&
\sum_{\ell,\ell^\prime}^{0,2}\sum_{i,j}^{k_i,k_j\le k_{\rm max}}
[P_\ell^{\rm theory}(k_i;\mathbf{p})-P_\ell^{\rm sim}(k_i)] \nonumber \\
&&(\mathbf{Cov}^{-1})_{\ell\ell^\prime}(k_i,k_j)
[P_{\ell^\prime}^{\rm theory}(k_j;\mathbf{p})-P_{\ell^\prime}^{\rm sim}(k_j)].
\nonumber \\
\end{eqnarray}
For the theoretical model, we adopt the 1-loop perturbative formulae derived in our previous work \citep{HKT19}. In order to correct the effects from small-scale physics on large-scale modes, we include the lowest-order counterterms given by $\alpha_\ell (\ell=0, 2)$ multiplied with $k^2$ times the linear power spectrum $P_\ell^{\rm linear}$  as suggested from the effective field theory \citep[e.g.,][]{Carrasco12}
\begin{equation}
P_\ell^{\rm theory} = P_\ell^{\rm 1-loop}+\alpha_\ell k^2 P_\ell^{\rm linear}(k).
\end{equation}
The set of free parameters $\mathbf{p}$ is the growth rate $f$ and two counterterms $\alpha_\ell$ and other cosmological parameters are fixed for simplicity. In the theoretical covariance of $P_\ell$  is again estimated from simulations with the survey volume $V=(500h^{-1}{\rm Mpc})^3$ and the number density $n\sim 1$[($h^{-1}$Mpc)$^{-3}]$ including a volume-size correction (eq.[\ref{eq:cov_sim2}]), which fully takes into account the mode coupling between different bins of $k$. Note that in our previous paper \citep{HKT19}, we assumed the Gaussian covariance with different volumes and number density for simplicity to estimate the impact of the growth rate measurement. We estimate the posterior distribution using a nested sampling algorithm {\tt multinest} \citep{Feroz09}, implemented in {\tt Monte Python} \citep{Audren13}. 

Figure \ref{fig:pkl_fit_z1} and \ref{fig:pkl_fit_z0} compare the monopole and quadrupole spectra from N-body simulations with the 1-loop perturbation theory at $z=1.02$ and 0. We adopt the best-fit values of $\alpha_\ell$ by fitting the spectrum out to $k=0.2h$/Mpc. Each panel shows the results of pre-rec (upper-left) and post-rec spectra with different $R_s=20h^{-1}$Mpc (upper-right), 10$h^{-1}$Mpc (lower-left) and 5$h^{-1}$Mpc (lower-right). We find that the fitting to the simulated spectrum is best for the post-rec spectra with $R_s=20h^{-1}$Mpc, while the post-rec spectra with $R_s=5h^{-1}$Mpc is the worst fitting. More quantitatively speaking, the minimum chi-squared values are 0.76 (pre-rec), 0.09 (post-rec with 20$h^{-1}$Mpc), 0.88 (post-rec with 10$h^{-1}$Mpc), and 3.6 (post-rec with 5$h^{-1}$Mpc) at $z=1.02$ and 8.9 (pre-rec), 0.7 (post-rec with 20$h^{-1}$Mpc), 4.7 (post-rec with 10$h^{-1}$Mpc), and 27 (post-rec with 5$h^{-1}$Mpc). Ref.~\cite{HKT19} showed that the reconstruction partially suppresses the nonlinearity of the gravitational growth and thereby the perturbation works at higher $k$. However, when $R_s$ is too small, the shift field estimated from the evolved density field becomes more nonlinear and thereby the perturbation does not work well. This is consistent with our previous work in reals-space matter clustering \citep{HKH17}. 

Figure \ref{fig:fz_err} shows the best-fit values of $f$ against the input value  $f_{\rm input}$ and its 1 sigma error. We find that the value of $f/f_{\rm input}$ is consistent with unity up to $k_{\rm max}\sim 0.3h$/Mpc at $z=1.02$ and $k_{\rm max}\sim 0.2h$/Mpc at $z=0$ for both pre-rec and post-rec spectra. Note that, since the 1-loop approximation does not work at such high $k_{\rm max}$, the input value of $f$ can be recovered by chance. It is found that the statistical error decreases after reconstruction over all range of $k$. For example, the error decrements of $f$ from the post-rec spectra relative to that from the pre-rec spectra at $z=1.02$ are 11\% ($R_s=20h^{-1}$Mpc), 20\% ($R_s=10h^{-1}$Mpc), and 17\% ($R_s=5h^{-1}$Mpc) at $k_{\rm max}=0.2h^{-1}$Mpc and 13\% ($R_s=20h^{-1}$Mpc), 33\% ($R_s=10h^{-1}$Mpc) and 33\% ($R_s=5h^{-1}$Mpc) at $k_{\rm max}=0.3h^{-1}$Mpc. The error decrements at $z=0$ are 18\% ($R_s=20h^{-1}$Mpc), 33\%  $R_s=10h^{-1}$Mpc), and 22\% ($R_s=5h^{-1}$Mpc) when $k_{\rm max}=0.2h^{-1}$Mpc. We find that the error of $f$ is significantly improved and the error improvement is almost maximized around $R_s\sim 10h^{-1}$Mpc where the covariance is almost diagonal by reconstruction. Table.~1 summarizes the minimum chi-squared values and the reduction of statistical errors for pre-rec and post-rec spectra with three different smoothing scales.  

\begin{table}[h!]
  \begin{center}
    \label{tab:table1}
    \begin{tabular}{c|c|c|c|c|c}
      &  & pre-rec & $20h^{-1}$Mpc& $10h^{-1}$Mpc & $5h^{-1}$Mpc \\
      \hline
    $\chi_{\rm min}^2$ & $z=1$ & 0.76 & 0.09 & 0.88 & 3.6 \\   
     & $z=0$ & 8.9 & 0.7 & 4.7 & 27 \\   
     \hline 
    error & $z=1$ & & & & \\
    reduction & $0.2h^{-1}$Mpc &- & 11$\%$ & 20(13)$\%$ & 17 $\%$ \\
      & $z=1$ & & & & \\
      & $0.3h^{-1}$Mpc &- & 13 $\%$ & 33(30)$\%$ & 33 $\%$ \\
      & $z=0$ & & & & \\
      & $0.2h^{-1}$Mpc &- & 18 $\%$ & 33(28)$\%$ & 22 $\%$ \\
      \hline 
    \end{tabular}
      \caption{A summary of the minimum chi-squared values and the reduction of statistical errors for pre-rec and post-rec spectra with three different smoothing scales $R_s$. For the reduction of statistical errors at $z=1$, the results with two different  $k_{\rm max}$ are shown. The error reduction for $R_s=10h^{-1}$Mpc shown in the bracket is the case with the super-sample covariance. 
      }
  \end{center}
\end{table}

\section{super-sample covariance}
\label{sec:ssc}

The super-sample covariance (SSC) comes from the mixing between the long-wavelength modes beyond the survey window and the short-wavelength modes inside the survey area. The response of the power spectrum to the change in background density $\delta_b$ is given as \citep{TakadaHu13,Li14}
\begin{equation}
\mathbf{Cov}^{\rm (SSC)}_{\ell\ell^\prime}=\sigma_{\rm b}^2\frac{\partial P_\ell(k)}{\partial \delta_{\rm b}}\frac{\partial P_{\ell^\prime}(k^\prime)}{\partial \delta_{\rm b}},
\end{equation}
where the variance of $\delta_{\rm b}$ in the survey window is defined as
\begin{equation}
\label{eq:sigb}
\sigma_{\rm b}^2=\frac{1}{V^2}\int \frac{\mathbf{dq}}{(2\pi)^3}
|\tilde{W}(\mathbf{q})|^2P_L(q),
\end{equation}with the Fourier transform of the survey mask field $W(\mathbf{x})$ given as $\tilde{W}(\mathbf{q})$.
The response of the multipole power spectrum to $\delta_{\rm b}$ is given by \citep{Li18}
\begin{equation}
\frac{\partial \ln P_\ell (k)}{\partial \delta_{\rm b}}= G_\ell + D_\ell\frac{d\ln k^3P_\ell(k)}{d\ln k},
\end{equation}
where the first term is the growth modulation by the background density, which is also known as beat coupling (BC) \citep{Hamilton06}, and the second term is the dilation effect that comes from the change of the local expansion rate depending on the background density  \citep{Li14}. Here we neglect the response of the background tide for simplicity.
The growth and dilation term for $\ell=0$ and 2 are given in Table 1 and 2 of \citep{Li18} as
\begin{eqnarray}
G_0&=&\frac{\frac{68}{21}(1+f)+\frac{164}{105}f^2+\frac{4}{15}f^3}{1+\frac{2}{3}f+\frac{1}{5}f^2}, \\
G_2&=&\frac{\frac{122}{21}f+\frac{656}{147}f^2+\frac{58}{63}f^3}{\frac{4}{3}f+\frac{4}{7}f^2}, 
\end{eqnarray}
and 
\begin{eqnarray}
D_0 &=& -\frac{\frac13(1+f)+\frac15f^2+\frac17f^3}{1+\frac23f+\frac15f^2}, \\
D_2&=&-\frac{\frac23f+\frac47f^2+\frac{10}{63}f^3}{\frac43f+\frac47f^2},
\end{eqnarray}
where the linear bias is set to be unity here.
The density fluctuation in a given survey is defined against the mean within the survey window rather than the global mean and thereby the normalization of the power spectrum is altered as \citep{TakadaHu13}
\begin{equation}
P_\ell^{\rm w}(k)=\frac{P_\ell(k)}{(1+\delta_{\rm b}^{\rm (z)})^2},
\end{equation}
where $\delta_{\rm b}^{\rm (z)}$ is the background density in redshift space and thereby the response is changed to 
\begin{equation}
\frac{\partial \ln P_\ell(k)}{\partial \delta_{\rm b}}\rightarrow
\frac{\partial \ln P_\ell^{\rm w}(k)}{\partial \delta_{\rm b}}\simeq
\frac{\partial \ln P_\ell(k)}{\partial \delta_{\rm b}}
- \left(2+\frac{2}{3}f\right).
\end{equation}
The non-Gaussian covariance is computed as the sum of the tree-level term (eq.[\ref{eq:cov_tree}]) and the SSC as
\begin{equation}
\mathbf{Cov}^{\rm (NG)}=\mathbf{Cov}^{\rm (tree)}+\mathbf{Cov}^{\rm (SSC)}.
\end{equation}

In section \ref{sec:nbody}, we numerically compute the covariance from the ensemble average over $N$-body simulations with the volume of $(500h^{-1}$Mpc$)^3$, however, the fluctuations beyond the boxsize are not taken into account. We compute the covariance including SSC by extracting subboxes with the volume of $(500h^{-1}$Mpc$)^3$ from the 8 realizations of large simulation boxes with the volume of $(4h^{-1}{\rm Gpc})^3$ containing $4096^3$ particles. The total number of subboxes becomes $8 \times (4h^{-1}{\rm Gpc}/500h^{-1}{\rm Mpc})^3 = 4096$. The mean density field is computed in each subbox and the shift field for reconstruction is computed from the smoothed density field using particle data in each subbox. The reconstructed density field is also computed in each subbox data including mass particles shifted from neighboring subboxes. Strictly speaking, the fluctuation beyond the large simulation boxsize  $4h^{-1}$Gpc is not included in the covariance, however, the SSC is dominated by the fluctuations below this size. \citep{Klypin18} also addressed the question of what box size is needed to model the large-scale structure and argued that modes larger than 1 $h^{-1}$Gpc do not contribute. For the purpose of comparison with the perturbation theory, however, we integrate $k$ from  $2\pi/(4h^{-1}$Gpc)  in the calculation of $\sigma_b$ (eq.[\ref{eq:sigb}]).

Figure \ref{fig:pkSN_ssc} shows the comparison of the S/N for pre-rec and post-rec spectra with $R_s=10h^{-1}$Mpc. We find that both S/N decrease when including SSC. Since the reconstruction is performed within the survey area, the bulk motion of super-sample modes cannot be corrected by the reconstruction. The post-rec spectra, however, have still higher S/N than the pre-rec spectra. The reconstruction improves S/N by 5\% ($k_{\rm max}=0.1h$/Mpc) and by 19\% ($k_{\rm max}=0.2h$/Mpc) at $z=1.02$ and by 14\% ($k_{\rm max}=0.1h$/Mpc) and by 40\% ($k_{\rm max}=0.2h$/Mpc) at $z=0$. The perturbation formulae also show the consistent results with the numerical one and they quantitatively agrees upto $k\sim 0.1h$/Mpc. 

Figure \ref{fig:fz_err_ssc} shows the impact on the growth rate measurements when including SSC. The input value of growth rate are again recovered upto $k_{\rm max}\sim 0.3h$/Mpc at $z=1.02$ and $k_{\rm max}\sim 0.2h$/Mpc at $z=0$ for both pre-rec and post-rec spectra. The improvements of the error by the reconstruction with $R_s=10h^{-1}$Mpc are 13\% when $k_{\rm max}=0.2h$/Mpc and 30\% when $k_{\rm max}=0.3h$/Mpc at $z=1.02$ and 28\% when $k_{\rm max}=0.2h$/Mpc at $z=0$, which are comparable to the improvement without SSC.

\section{Summary and conclusions}
\label{sec:summary}

We investigated the covariance of the redshift-space matter power spectra after a standard density-field reconstruction that is commonly used in the BAO analysis. We derived the perturbative formula of the covariance of the multipole components of the power spectra at tree level. We find that the positive off-diagonal components of the covariance from the tree-level trispectra decrease after the reconstruction and have negative values at the smoothing scale of the shift field $R_s$ less than $\sim 10h^{-1}$Mpc. We also computed the covariance of the multipole power spectra directly from a large set of $N$-body simulations. We find the significant decrease of the off-diagonal components and the behavior is consistent with the perturbation theory. In consequence, the information content of the post-rec power spectra evaluated with the signal-to-noise ratio (S/N) of their monopole and quadrupole components significantly increase compared to the pre-rec power spectra. Interestingly, the S/N of the reconstructed spectra with $R_s$ less than 10$h^{-1}$Mpc exceeds to that of the linear spectrum with the Gaussian covariance, which comes from the negative off-diagonal components of the covariance matrix. The enhancement of the S/N is more significant at later times. We also studied the super-sample covariance effect both from perturbative and numerical approaches. We find that the S/N reduces even after the reconstruction because the reconstruction performs within the survey area and thus the bulk motion of the super-sample modes cannot be corrected by reconstruction. Even when the SSC is included, the post-rec spectra still have higher S/N than the pre-rec spectra.

We find that the tree-level perturbative approach is limited to describe the simulated covariance at $k\le 0.1h$/Mpc. This indicates that higher-order mode coupling needs to be taken into account to describe the covariance more accurately. There are several works to describe mode couplings at higher $k$ based on the effective field theory \citep{Bertolini16}, the response approach \citep{Barreira17}, and also semi-analytical models \citep{Neyrinck11,Mohammed14,Mohammed17}. It may be interesting to apply these methods to describe the covariance of reconstructed spectra.

Recovery of cosmological information in the two-point statistics makes the cosmological analysis simpler. We demonstrated that the reconstruction significantly reduced the error of growth rate inferred from the redshift-space power spectrum. So far the reconstruction has been mainly applied to the BAO analysis due to the lack of theoretical understandings of the reconstructed spectrum. Since it is found that the error of the full shape of the power spectrum is improved, it is interesting to investigate how the other cosmological parameters are improved by using the information of the full shape of power spectra after reconstruction. We also have to take into account the galaxy bias and the shot noise as well as various observational effects such as survey geometry to apply the actual observational data \citep[e.g.,][]{Wadekar19}. The shot noise increases the statistical uncertainties in the power spectrum(eq.[\ref{eq:cov_ga}]) and also in the shift field for reconstruction particularly when the smoothing scale is small. We leave this for future work. 

\begin{acknowledgments}
We thank an anonymous referee for useful comments.
This work is supported by MEXT/JSPS KAKENHI Grant Numbers
JP16K17684 (CH), JP18H04348 (CH), JP17H01131 (RT), and 20H04723 (RT). KK is supported by the UK STFC grant ST/S000550/1, and the European Research Council under the European Union's Horizon 2020 programme (grant agreement No.646702 "CosTesGrav").
Numerical computations were in part carried out on Cray XC30 and XC50 at Centre for Computational Astrophysics, National Astronomical Observatory of Japan.
\end{acknowledgments}

\bibliographystyle{apsrev}
\bibliography{ref}

\begin{thebibliography}{87}
\expandafter\ifx\csname natexlab\endcsname\relax\def\natexlab#1{#1}\fi
\expandafter\ifx\csname bibnamefont\endcsname\relax
  \def\bibnamefont#1{#1}\fi
\expandafter\ifx\csname bibfnamefont\endcsname\relax
  \def\bibfnamefont#1{#1}\fi
\expandafter\ifx\csname citenamefont\endcsname\relax
  \def\citenamefont#1{#1}\fi
\expandafter\ifx\csname url\endcsname\relax
  \def\url#1{\texttt{#1}}\fi
\expandafter\ifx\csname urlprefix\endcsname\relax\def\urlprefix{URL }\fi
\providecommand{\bibinfo}[2]{#2}
\providecommand{\eprint}[2][]{\url{#2}}

\bibitem[{\citenamefont{{Amendola} et~al.}(2005)\citenamefont{{Amendola},
  {Quercellini}, and {Giallongo}}}]{Amendola05}
\bibinfo{author}{\bibfnamefont{L.}~\bibnamefont{{Amendola}}},
  \bibinfo{author}{\bibfnamefont{C.}~\bibnamefont{{Quercellini}}},
  \bibnamefont{and}
  \bibinfo{author}{\bibfnamefont{E.}~\bibnamefont{{Giallongo}}},
  \bibinfo{journal}{\mnras} \textbf{\bibinfo{volume}{357}},
  \bibinfo{pages}{429} (\bibinfo{year}{2005}), \eprint{astro-ph/0404599}.

\bibitem[{\citenamefont{{Eisenstein} et~al.}(1998)\citenamefont{{Eisenstein},
  {Hu}, and {Tegmark}}}]{Eisenstein98}
\bibinfo{author}{\bibfnamefont{D.~J.} \bibnamefont{{Eisenstein}}},
  \bibinfo{author}{\bibfnamefont{W.}~\bibnamefont{{Hu}}}, \bibnamefont{and}
  \bibinfo{author}{\bibfnamefont{M.}~\bibnamefont{{Tegmark}}},
  \bibinfo{journal}{\apjl} \textbf{\bibinfo{volume}{504}}, \bibinfo{pages}{L57}
  (\bibinfo{year}{1998}), \eprint{astro-ph/9805239}.

\bibitem[{\citenamefont{{Meiksin} et~al.}(1999)\citenamefont{{Meiksin},
  {White}, and {Peacock}}}]{Meiksin99}
\bibinfo{author}{\bibfnamefont{A.}~\bibnamefont{{Meiksin}}},
  \bibinfo{author}{\bibfnamefont{M.}~\bibnamefont{{White}}}, \bibnamefont{and}
  \bibinfo{author}{\bibfnamefont{J.~A.} \bibnamefont{{Peacock}}},
  \bibinfo{journal}{\mnras} \textbf{\bibinfo{volume}{304}},
  \bibinfo{pages}{851} (\bibinfo{year}{1999}), \eprint{astro-ph/9812214}.

\bibitem[{\citenamefont{{Blake} and {Glazebrook}}(2003)}]{BlakeGlazebrook03}
\bibinfo{author}{\bibfnamefont{C.}~\bibnamefont{{Blake}}} \bibnamefont{and}
  \bibinfo{author}{\bibfnamefont{K.}~\bibnamefont{{Glazebrook}}},
  \bibinfo{journal}{\apj} \textbf{\bibinfo{volume}{594}}, \bibinfo{pages}{665}
  (\bibinfo{year}{2003}), \eprint{astro-ph/0301632}.

\bibitem[{\citenamefont{{Hu} and {Haiman}}(2003)}]{HuHaiman03}
\bibinfo{author}{\bibfnamefont{W.}~\bibnamefont{{Hu}}} \bibnamefont{and}
  \bibinfo{author}{\bibfnamefont{Z.}~\bibnamefont{{Haiman}}},
  \bibinfo{journal}{\prd} \textbf{\bibinfo{volume}{68}}, \bibinfo{eid}{063004}
  (\bibinfo{year}{2003}), \eprint{astro-ph/0306053}.

\bibitem[{\citenamefont{{Matsubara}}(2004)}]{Matsubara04}
\bibinfo{author}{\bibfnamefont{T.}~\bibnamefont{{Matsubara}}},
  \bibinfo{journal}{\apj} \textbf{\bibinfo{volume}{615}}, \bibinfo{pages}{573}
  (\bibinfo{year}{2004}), \eprint{astro-ph/0408349}.

\bibitem[{\citenamefont{{Angulo} et~al.}(2005)\citenamefont{{Angulo}, {Baugh},
  {Frenk}, {Bower}, {Jenkins}, and {Morris}}}]{Angulo05}
\bibinfo{author}{\bibfnamefont{R.~E.} \bibnamefont{{Angulo}}},
  \bibinfo{author}{\bibfnamefont{C.~M.} \bibnamefont{{Baugh}}},
  \bibinfo{author}{\bibfnamefont{C.~S.} \bibnamefont{{Frenk}}},
  \bibinfo{author}{\bibfnamefont{R.~G.} \bibnamefont{{Bower}}},
  \bibinfo{author}{\bibfnamefont{A.}~\bibnamefont{{Jenkins}}},
  \bibnamefont{and} \bibinfo{author}{\bibfnamefont{S.~L.}
  \bibnamefont{{Morris}}}, \bibinfo{journal}{\mnras}
  \textbf{\bibinfo{volume}{362}}, \bibinfo{pages}{L25} (\bibinfo{year}{2005}),
  \eprint{astro-ph/0504456}.

\bibitem[{\citenamefont{{Seo} and {Eisenstein}}(2005)}]{SeoEisenstein05}
\bibinfo{author}{\bibfnamefont{H.-J.} \bibnamefont{{Seo}}} \bibnamefont{and}
  \bibinfo{author}{\bibfnamefont{D.~J.} \bibnamefont{{Eisenstein}}},
  \bibinfo{journal}{\apj} \textbf{\bibinfo{volume}{633}}, \bibinfo{pages}{575}
  (\bibinfo{year}{2005}), \eprint{astro-ph/0507338}.

\bibitem[{\citenamefont{{White}}(2005)}]{White05}
\bibinfo{author}{\bibfnamefont{M.}~\bibnamefont{{White}}},
  \bibinfo{journal}{Astroparticle Physics} \textbf{\bibinfo{volume}{24}},
  \bibinfo{pages}{334} (\bibinfo{year}{2005}), \eprint{astro-ph/0507307}.

\bibitem[{\citenamefont{{Eisenstein} et~al.}(2005)\citenamefont{{Eisenstein},
  {Zehavi}, {Hogg}, {Scoccimarro}, {Blanton}, {Nichol}, {Scranton}, {Seo},
  {Tegmark}, {Zheng} et~al.}}]{Eisenstein05}
\bibinfo{author}{\bibfnamefont{D.~J.} \bibnamefont{{Eisenstein}}},
  \bibinfo{author}{\bibfnamefont{I.}~\bibnamefont{{Zehavi}}},
  \bibinfo{author}{\bibfnamefont{D.~W.} \bibnamefont{{Hogg}}},
  \bibinfo{author}{\bibfnamefont{R.}~\bibnamefont{{Scoccimarro}}},
  \bibinfo{author}{\bibfnamefont{M.~R.} \bibnamefont{{Blanton}}},
  \bibinfo{author}{\bibfnamefont{R.~C.} \bibnamefont{{Nichol}}},
  \bibinfo{author}{\bibfnamefont{R.}~\bibnamefont{{Scranton}}},
  \bibinfo{author}{\bibfnamefont{H.-J.} \bibnamefont{{Seo}}},
  \bibinfo{author}{\bibfnamefont{M.}~\bibnamefont{{Tegmark}}},
  \bibinfo{author}{\bibfnamefont{Z.}~\bibnamefont{{Zheng}}},
  \bibnamefont{et~al.}, \bibinfo{journal}{\apj} \textbf{\bibinfo{volume}{633}},
  \bibinfo{pages}{560} (\bibinfo{year}{2005}), \eprint{astro-ph/0501171}.

\bibitem[{\citenamefont{{Cole} et~al.}(2005)\citenamefont{{Cole}, {Percival},
  {Peacock}, {Norberg}, {Baugh}, {Frenk}, {Baldry}, {Bland-Hawthorn},
  {Bridges}, {Cannon} et~al.}}]{Cole05}
\bibinfo{author}{\bibfnamefont{S.}~\bibnamefont{{Cole}}},
  \bibinfo{author}{\bibfnamefont{W.~J.} \bibnamefont{{Percival}}},
  \bibinfo{author}{\bibfnamefont{J.~A.} \bibnamefont{{Peacock}}},
  \bibinfo{author}{\bibfnamefont{P.}~\bibnamefont{{Norberg}}},
  \bibinfo{author}{\bibfnamefont{C.~M.} \bibnamefont{{Baugh}}},
  \bibinfo{author}{\bibfnamefont{C.~S.} \bibnamefont{{Frenk}}},
  \bibinfo{author}{\bibfnamefont{I.}~\bibnamefont{{Baldry}}},
  \bibinfo{author}{\bibfnamefont{J.}~\bibnamefont{{Bland-Hawthorn}}},
  \bibinfo{author}{\bibfnamefont{T.}~\bibnamefont{{Bridges}}},
  \bibinfo{author}{\bibfnamefont{R.}~\bibnamefont{{Cannon}}},
  \bibnamefont{et~al.}, \bibinfo{journal}{\mnras}
  \textbf{\bibinfo{volume}{362}}, \bibinfo{pages}{505} (\bibinfo{year}{2005}),
  \eprint{astro-ph/0501174}.

\bibitem[{\citenamefont{{Padmanabhan} et~al.}(2007)\citenamefont{{Padmanabhan},
  {Schlegel}, {Seljak}, {Makarov}, {Bahcall}, {Blanton}, {Brinkmann},
  {Eisenstein}, {Finkbeiner}, {Gunn} et~al.}}]{Padmanabhan07}
\bibinfo{author}{\bibfnamefont{N.}~\bibnamefont{{Padmanabhan}}},
  \bibinfo{author}{\bibfnamefont{D.~J.} \bibnamefont{{Schlegel}}},
  \bibinfo{author}{\bibfnamefont{U.}~\bibnamefont{{Seljak}}},
  \bibinfo{author}{\bibfnamefont{A.}~\bibnamefont{{Makarov}}},
  \bibinfo{author}{\bibfnamefont{N.~A.} \bibnamefont{{Bahcall}}},
  \bibinfo{author}{\bibfnamefont{M.~R.} \bibnamefont{{Blanton}}},
  \bibinfo{author}{\bibfnamefont{J.}~\bibnamefont{{Brinkmann}}},
  \bibinfo{author}{\bibfnamefont{D.~J.} \bibnamefont{{Eisenstein}}},
  \bibinfo{author}{\bibfnamefont{D.~P.} \bibnamefont{{Finkbeiner}}},
  \bibinfo{author}{\bibfnamefont{J.~E.} \bibnamefont{{Gunn}}},
  \bibnamefont{et~al.}, \bibinfo{journal}{\mnras}
  \textbf{\bibinfo{volume}{378}}, \bibinfo{pages}{852} (\bibinfo{year}{2007}),
  \eprint{astro-ph/0605302}.

\bibitem[{\citenamefont{{Percival} et~al.}(2007)\citenamefont{{Percival},
  {Nichol}, {Eisenstein}, {Frieman}, {Fukugita}, {Loveday}, {Pope},
  {Schneider}, {Szalay}, {Tegmark} et~al.}}]{Percival07}
\bibinfo{author}{\bibfnamefont{W.~J.} \bibnamefont{{Percival}}},
  \bibinfo{author}{\bibfnamefont{R.~C.} \bibnamefont{{Nichol}}},
  \bibinfo{author}{\bibfnamefont{D.~J.} \bibnamefont{{Eisenstein}}},
  \bibinfo{author}{\bibfnamefont{J.~A.} \bibnamefont{{Frieman}}},
  \bibinfo{author}{\bibfnamefont{M.}~\bibnamefont{{Fukugita}}},
  \bibinfo{author}{\bibfnamefont{J.}~\bibnamefont{{Loveday}}},
  \bibinfo{author}{\bibfnamefont{A.~C.} \bibnamefont{{Pope}}},
  \bibinfo{author}{\bibfnamefont{D.~P.} \bibnamefont{{Schneider}}},
  \bibinfo{author}{\bibfnamefont{A.~S.} \bibnamefont{{Szalay}}},
  \bibinfo{author}{\bibfnamefont{M.}~\bibnamefont{{Tegmark}}},
  \bibnamefont{et~al.}, \bibinfo{journal}{\apj} \textbf{\bibinfo{volume}{657}},
  \bibinfo{pages}{645} (\bibinfo{year}{2007}), \eprint{astro-ph/0608636}.

\bibitem[{\citenamefont{{Eisenstein}
  et~al.}(2007{\natexlab{a}})\citenamefont{{Eisenstein}, {Seo}, and
  {White}}}]{Eisenstein07a}
\bibinfo{author}{\bibfnamefont{D.~J.} \bibnamefont{{Eisenstein}}},
  \bibinfo{author}{\bibfnamefont{H.-J.} \bibnamefont{{Seo}}}, \bibnamefont{and}
  \bibinfo{author}{\bibfnamefont{M.}~\bibnamefont{{White}}},
  \bibinfo{journal}{\apj} \textbf{\bibinfo{volume}{664}}, \bibinfo{pages}{660}
  (\bibinfo{year}{2007}{\natexlab{a}}), \eprint{astro-ph/0604361}.

\bibitem[{\citenamefont{{Huff} et~al.}(2007)\citenamefont{{Huff}, {Schulz},
  {White}, {Schlegel}, and {Warren}}}]{Huff07}
\bibinfo{author}{\bibfnamefont{E.}~\bibnamefont{{Huff}}},
  \bibinfo{author}{\bibfnamefont{A.~E.} \bibnamefont{{Schulz}}},
  \bibinfo{author}{\bibfnamefont{M.}~\bibnamefont{{White}}},
  \bibinfo{author}{\bibfnamefont{D.~J.} \bibnamefont{{Schlegel}}},
  \bibnamefont{and} \bibinfo{author}{\bibfnamefont{M.~S.}
  \bibnamefont{{Warren}}}, \bibinfo{journal}{Astroparticle Physics}
  \textbf{\bibinfo{volume}{26}}, \bibinfo{pages}{351} (\bibinfo{year}{2007}),
  \eprint{astro-ph/0607061}.

\bibitem[{\citenamefont{{Angulo} et~al.}(2008)\citenamefont{{Angulo}, {Baugh},
  {Frenk}, and {Lacey}}}]{Angulo08}
\bibinfo{author}{\bibfnamefont{R.~E.} \bibnamefont{{Angulo}}},
  \bibinfo{author}{\bibfnamefont{C.~M.} \bibnamefont{{Baugh}}},
  \bibinfo{author}{\bibfnamefont{C.~S.} \bibnamefont{{Frenk}}},
  \bibnamefont{and} \bibinfo{author}{\bibfnamefont{C.~G.}
  \bibnamefont{{Lacey}}}, \bibinfo{journal}{\mnras}
  \textbf{\bibinfo{volume}{383}}, \bibinfo{pages}{755} (\bibinfo{year}{2008}),
  \eprint{astro-ph/0702543}.

\bibitem[{\citenamefont{{Okumura} et~al.}(2008)\citenamefont{{Okumura},
  {Matsubara}, {Eisenstein}, {Kayo}, {Hikage}, {Szalay}, and
  {Schneider}}}]{Okumura08}
\bibinfo{author}{\bibfnamefont{T.}~\bibnamefont{{Okumura}}},
  \bibinfo{author}{\bibfnamefont{T.}~\bibnamefont{{Matsubara}}},
  \bibinfo{author}{\bibfnamefont{D.~J.} \bibnamefont{{Eisenstein}}},
  \bibinfo{author}{\bibfnamefont{I.}~\bibnamefont{{Kayo}}},
  \bibinfo{author}{\bibfnamefont{C.}~\bibnamefont{{Hikage}}},
  \bibinfo{author}{\bibfnamefont{A.~S.} \bibnamefont{{Szalay}}},
  \bibnamefont{and} \bibinfo{author}{\bibfnamefont{D.~P.}
  \bibnamefont{{Schneider}}}, \bibinfo{journal}{\apj}
  \textbf{\bibinfo{volume}{676}}, \bibinfo{pages}{889} (\bibinfo{year}{2008}),
  \eprint{0711.3640}.

\bibitem[{\citenamefont{{Xu} et~al.}(2013)\citenamefont{{Xu}, {Cuesta},
  {Padmanabhan}, {Eisenstein}, and {McBride}}}]{Xu13}
\bibinfo{author}{\bibfnamefont{X.}~\bibnamefont{{Xu}}},
  \bibinfo{author}{\bibfnamefont{A.~J.} \bibnamefont{{Cuesta}}},
  \bibinfo{author}{\bibfnamefont{N.}~\bibnamefont{{Padmanabhan}}},
  \bibinfo{author}{\bibfnamefont{D.~J.} \bibnamefont{{Eisenstein}}},
  \bibnamefont{and} \bibinfo{author}{\bibfnamefont{C.~K.}
  \bibnamefont{{McBride}}}, \bibinfo{journal}{\mnras}
  \textbf{\bibinfo{volume}{431}}, \bibinfo{pages}{2834} (\bibinfo{year}{2013}),
  \eprint{1206.6732}.

\bibitem[{\citenamefont{{Anderson} et~al.}(2014)\citenamefont{{Anderson},
  {Aubourg}, {Bailey}, {Beutler}, {Bhardwaj}, {Blanton}, {Bolton}, {Brinkmann},
  {Brownstein}, {Burden} et~al.}}]{Anderson14}
\bibinfo{author}{\bibfnamefont{L.}~\bibnamefont{{Anderson}}},
  \bibinfo{author}{\bibfnamefont{{\'E}.}~\bibnamefont{{Aubourg}}},
  \bibinfo{author}{\bibfnamefont{S.}~\bibnamefont{{Bailey}}},
  \bibinfo{author}{\bibfnamefont{F.}~\bibnamefont{{Beutler}}},
  \bibinfo{author}{\bibfnamefont{V.}~\bibnamefont{{Bhardwaj}}},
  \bibinfo{author}{\bibfnamefont{M.}~\bibnamefont{{Blanton}}},
  \bibinfo{author}{\bibfnamefont{A.~S.} \bibnamefont{{Bolton}}},
  \bibinfo{author}{\bibfnamefont{J.}~\bibnamefont{{Brinkmann}}},
  \bibinfo{author}{\bibfnamefont{J.~R.} \bibnamefont{{Brownstein}}},
  \bibinfo{author}{\bibfnamefont{A.}~\bibnamefont{{Burden}}},
  \bibnamefont{et~al.}, \bibinfo{journal}{\mnras}
  \textbf{\bibinfo{volume}{441}}, \bibinfo{pages}{24} (\bibinfo{year}{2014}),
  \eprint{1312.4877}.

\bibitem[{\citenamefont{{Tojeiro} et~al.}(2014)\citenamefont{{Tojeiro}, {Ross},
  {Burden}, {Samushia}, {Manera}, {Percival}, {Beutler}, {Brinkmann},
  {Brownstein}, {Cuesta} et~al.}}]{Tojeiro14}
\bibinfo{author}{\bibfnamefont{R.}~\bibnamefont{{Tojeiro}}},
  \bibinfo{author}{\bibfnamefont{A.~J.} \bibnamefont{{Ross}}},
  \bibinfo{author}{\bibfnamefont{A.}~\bibnamefont{{Burden}}},
  \bibinfo{author}{\bibfnamefont{L.}~\bibnamefont{{Samushia}}},
  \bibinfo{author}{\bibfnamefont{M.}~\bibnamefont{{Manera}}},
  \bibinfo{author}{\bibfnamefont{W.~J.} \bibnamefont{{Percival}}},
  \bibinfo{author}{\bibfnamefont{F.}~\bibnamefont{{Beutler}}},
  \bibinfo{author}{\bibfnamefont{J.}~\bibnamefont{{Brinkmann}}},
  \bibinfo{author}{\bibfnamefont{J.~R.} \bibnamefont{{Brownstein}}},
  \bibinfo{author}{\bibfnamefont{A.~J.} \bibnamefont{{Cuesta}}},
  \bibnamefont{et~al.}, \bibinfo{journal}{\mnras}
  \textbf{\bibinfo{volume}{440}}, \bibinfo{pages}{2222} (\bibinfo{year}{2014}),
  \eprint{1401.1768}.

\bibitem[{\citenamefont{{Kazin} et~al.}(2014)\citenamefont{{Kazin}, {Koda},
  {Blake}, {Padmanabhan}, {Brough}, {Colless}, {Contreras}, {Couch}, {Croom},
  {Croton} et~al.}}]{Kazin14}
\bibinfo{author}{\bibfnamefont{E.~A.} \bibnamefont{{Kazin}}},
  \bibinfo{author}{\bibfnamefont{J.}~\bibnamefont{{Koda}}},
  \bibinfo{author}{\bibfnamefont{C.}~\bibnamefont{{Blake}}},
  \bibinfo{author}{\bibfnamefont{N.}~\bibnamefont{{Padmanabhan}}},
  \bibinfo{author}{\bibfnamefont{S.}~\bibnamefont{{Brough}}},
  \bibinfo{author}{\bibfnamefont{M.}~\bibnamefont{{Colless}}},
  \bibinfo{author}{\bibfnamefont{C.}~\bibnamefont{{Contreras}}},
  \bibinfo{author}{\bibfnamefont{W.}~\bibnamefont{{Couch}}},
  \bibinfo{author}{\bibfnamefont{S.}~\bibnamefont{{Croom}}},
  \bibinfo{author}{\bibfnamefont{D.~J.} \bibnamefont{{Croton}}},
  \bibnamefont{et~al.}, \bibinfo{journal}{\mnras}
  \textbf{\bibinfo{volume}{441}}, \bibinfo{pages}{3524} (\bibinfo{year}{2014}),
  \eprint{1401.0358}.

\bibitem[{\citenamefont{{Ross} et~al.}(2015)\citenamefont{{Ross}, {Samushia},
  {Howlett}, {Percival}, {Burden}, and {Manera}}}]{Ross15}
\bibinfo{author}{\bibfnamefont{A.~J.} \bibnamefont{{Ross}}},
  \bibinfo{author}{\bibfnamefont{L.}~\bibnamefont{{Samushia}}},
  \bibinfo{author}{\bibfnamefont{C.}~\bibnamefont{{Howlett}}},
  \bibinfo{author}{\bibfnamefont{W.~J.} \bibnamefont{{Percival}}},
  \bibinfo{author}{\bibfnamefont{A.}~\bibnamefont{{Burden}}}, \bibnamefont{and}
  \bibinfo{author}{\bibfnamefont{M.}~\bibnamefont{{Manera}}},
  \bibinfo{journal}{\mnras} \textbf{\bibinfo{volume}{449}},
  \bibinfo{pages}{835} (\bibinfo{year}{2015}), \eprint{1409.3242}.

\bibitem[{\citenamefont{{Alam} et~al.}(2016)\citenamefont{{Alam}, {Ata},
  {Bailey}, {Beutler}, {Bizyaev}, {Blazek}, {Bolton}, {Brownstein}, {Burden},
  {Chuang} et~al.}}]{Alam16}
\bibinfo{author}{\bibfnamefont{S.}~\bibnamefont{{Alam}}},
  \bibinfo{author}{\bibfnamefont{M.}~\bibnamefont{{Ata}}},
  \bibinfo{author}{\bibfnamefont{S.}~\bibnamefont{{Bailey}}},
  \bibinfo{author}{\bibfnamefont{F.}~\bibnamefont{{Beutler}}},
  \bibinfo{author}{\bibfnamefont{D.}~\bibnamefont{{Bizyaev}}},
  \bibinfo{author}{\bibfnamefont{J.~A.} \bibnamefont{{Blazek}}},
  \bibinfo{author}{\bibfnamefont{A.~S.} \bibnamefont{{Bolton}}},
  \bibinfo{author}{\bibfnamefont{J.~R.} \bibnamefont{{Brownstein}}},
  \bibinfo{author}{\bibfnamefont{A.}~\bibnamefont{{Burden}}},
  \bibinfo{author}{\bibfnamefont{C.-H.} \bibnamefont{{Chuang}}},
  \bibnamefont{et~al.}, \bibinfo{journal}{ArXiv e-prints}
  (\bibinfo{year}{2016}), \eprint{1607.03155}.

\bibitem[{\citenamefont{{Beutler} et~al.}(2017)\citenamefont{{Beutler}, {Seo},
  {Ross}, {McDonald}, {Saito}, {Bolton}, {Brownstein}, {Chuang}, {Cuesta},
  {Eisenstein} et~al.}}]{Beutler17}
\bibinfo{author}{\bibfnamefont{F.}~\bibnamefont{{Beutler}}},
  \bibinfo{author}{\bibfnamefont{H.-J.} \bibnamefont{{Seo}}},
  \bibinfo{author}{\bibfnamefont{A.~J.} \bibnamefont{{Ross}}},
  \bibinfo{author}{\bibfnamefont{P.}~\bibnamefont{{McDonald}}},
  \bibinfo{author}{\bibfnamefont{S.}~\bibnamefont{{Saito}}},
  \bibinfo{author}{\bibfnamefont{A.~S.} \bibnamefont{{Bolton}}},
  \bibinfo{author}{\bibfnamefont{J.~R.} \bibnamefont{{Brownstein}}},
  \bibinfo{author}{\bibfnamefont{C.-H.} \bibnamefont{{Chuang}}},
  \bibinfo{author}{\bibfnamefont{A.~J.} \bibnamefont{{Cuesta}}},
  \bibinfo{author}{\bibfnamefont{D.~J.} \bibnamefont{{Eisenstein}}},
  \bibnamefont{et~al.}, \bibinfo{journal}{\mnras}
  \textbf{\bibinfo{volume}{464}}, \bibinfo{pages}{3409} (\bibinfo{year}{2017}),
  \eprint{1607.03149}.

\bibitem[{\citenamefont{{Guzzo} et~al.}(2008)\citenamefont{{Guzzo},
  {Pierleoni}, {Meneux}, {Branchini}, {Le F{\`e}vre}, {Marinoni}, {Garilli},
  {Blaizot}, {De Lucia}, {Pollo} et~al.}}]{Guzzo08}
\bibinfo{author}{\bibfnamefont{L.}~\bibnamefont{{Guzzo}}},
  \bibinfo{author}{\bibfnamefont{M.}~\bibnamefont{{Pierleoni}}},
  \bibinfo{author}{\bibfnamefont{B.}~\bibnamefont{{Meneux}}},
  \bibinfo{author}{\bibfnamefont{E.}~\bibnamefont{{Branchini}}},
  \bibinfo{author}{\bibfnamefont{O.}~\bibnamefont{{Le F{\`e}vre}}},
  \bibinfo{author}{\bibfnamefont{C.}~\bibnamefont{{Marinoni}}},
  \bibinfo{author}{\bibfnamefont{B.}~\bibnamefont{{Garilli}}},
  \bibinfo{author}{\bibfnamefont{J.}~\bibnamefont{{Blaizot}}},
  \bibinfo{author}{\bibfnamefont{G.}~\bibnamefont{{De Lucia}}},
  \bibinfo{author}{\bibfnamefont{A.}~\bibnamefont{{Pollo}}},
  \bibnamefont{et~al.}, \bibinfo{journal}{\nat} \textbf{\bibinfo{volume}{451}},
  \bibinfo{pages}{541} (\bibinfo{year}{2008}), \eprint{0802.1944}.

\bibitem[{\citenamefont{{Yamamoto} et~al.}(2008)\citenamefont{{Yamamoto},
  {Sato}, and {H{\"u}tsi}}}]{Yamamoto08}
\bibinfo{author}{\bibfnamefont{K.}~\bibnamefont{{Yamamoto}}},
  \bibinfo{author}{\bibfnamefont{T.}~\bibnamefont{{Sato}}}, \bibnamefont{and}
  \bibinfo{author}{\bibfnamefont{G.}~\bibnamefont{{H{\"u}tsi}}},
  \bibinfo{journal}{Progress of Theoretical Physics}
  \textbf{\bibinfo{volume}{120}}, \bibinfo{pages}{609} (\bibinfo{year}{2008}),
  \eprint{0805.4789}.

\bibitem[{\citenamefont{{Reid} et~al.}(2012)\citenamefont{{Reid}, {Samushia},
  {White}, {Percival}, {Manera}, {Padmanabhan}, {Ross}, {S{\'a}nchez},
  {Bailey}, {Bizyaev} et~al.}}]{Reid12}
\bibinfo{author}{\bibfnamefont{B.~A.} \bibnamefont{{Reid}}},
  \bibinfo{author}{\bibfnamefont{L.}~\bibnamefont{{Samushia}}},
  \bibinfo{author}{\bibfnamefont{M.}~\bibnamefont{{White}}},
  \bibinfo{author}{\bibfnamefont{W.~J.} \bibnamefont{{Percival}}},
  \bibinfo{author}{\bibfnamefont{M.}~\bibnamefont{{Manera}}},
  \bibinfo{author}{\bibfnamefont{N.}~\bibnamefont{{Padmanabhan}}},
  \bibinfo{author}{\bibfnamefont{A.~J.} \bibnamefont{{Ross}}},
  \bibinfo{author}{\bibfnamefont{A.~G.} \bibnamefont{{S{\'a}nchez}}},
  \bibinfo{author}{\bibfnamefont{S.}~\bibnamefont{{Bailey}}},
  \bibinfo{author}{\bibfnamefont{D.}~\bibnamefont{{Bizyaev}}},
  \bibnamefont{et~al.}, \bibinfo{journal}{\mnras}
  \textbf{\bibinfo{volume}{426}}, \bibinfo{pages}{2719} (\bibinfo{year}{2012}),
  \eprint{1203.6641}.

\bibitem[{\citenamefont{{Beutler} et~al.}(2014)\citenamefont{{Beutler},
  {Saito}, {Seo}, {Brinkmann}, {Dawson}, {Eisenstein}, {Font-Ribera}, {Ho},
  {McBride}, {Montesano} et~al.}}]{Beutler13}
\bibinfo{author}{\bibfnamefont{F.}~\bibnamefont{{Beutler}}},
  \bibinfo{author}{\bibfnamefont{S.}~\bibnamefont{{Saito}}},
  \bibinfo{author}{\bibfnamefont{H.-J.} \bibnamefont{{Seo}}},
  \bibinfo{author}{\bibfnamefont{J.}~\bibnamefont{{Brinkmann}}},
  \bibinfo{author}{\bibfnamefont{K.~S.} \bibnamefont{{Dawson}}},
  \bibinfo{author}{\bibfnamefont{D.~J.} \bibnamefont{{Eisenstein}}},
  \bibinfo{author}{\bibfnamefont{A.}~\bibnamefont{{Font-Ribera}}},
  \bibinfo{author}{\bibfnamefont{S.}~\bibnamefont{{Ho}}},
  \bibinfo{author}{\bibfnamefont{C.~K.} \bibnamefont{{McBride}}},
  \bibinfo{author}{\bibfnamefont{F.}~\bibnamefont{{Montesano}}},
  \bibnamefont{et~al.}, \bibinfo{journal}{\mnras}
  \textbf{\bibinfo{volume}{443}}, \bibinfo{pages}{1065} (\bibinfo{year}{2014}),
  \eprint{1312.4611}.

\bibitem[{\citenamefont{{Samushia} et~al.}(2014)\citenamefont{{Samushia},
  {Reid}, {White}, {Percival}, {Cuesta}, {Zhao}, {Ross}, {Manera}, {Aubourg},
  {Beutler} et~al.}}]{Samushia13}
\bibinfo{author}{\bibfnamefont{L.}~\bibnamefont{{Samushia}}},
  \bibinfo{author}{\bibfnamefont{B.~A.} \bibnamefont{{Reid}}},
  \bibinfo{author}{\bibfnamefont{M.}~\bibnamefont{{White}}},
  \bibinfo{author}{\bibfnamefont{W.~J.} \bibnamefont{{Percival}}},
  \bibinfo{author}{\bibfnamefont{A.~J.} \bibnamefont{{Cuesta}}},
  \bibinfo{author}{\bibfnamefont{G.-B.} \bibnamefont{{Zhao}}},
  \bibinfo{author}{\bibfnamefont{A.~J.} \bibnamefont{{Ross}}},
  \bibinfo{author}{\bibfnamefont{M.}~\bibnamefont{{Manera}}},
  \bibinfo{author}{\bibfnamefont{{\'E}.}~\bibnamefont{{Aubourg}}},
  \bibinfo{author}{\bibfnamefont{F.}~\bibnamefont{{Beutler}}},
  \bibnamefont{et~al.}, \bibinfo{journal}{\mnras}
  \textbf{\bibinfo{volume}{439}}, \bibinfo{pages}{3504} (\bibinfo{year}{2014}),
  \eprint{1312.4899}.

\bibitem[{\citenamefont{{Oka} et~al.}(2014)\citenamefont{{Oka}, {Saito},
  {Nishimichi}, {Taruya}, and {Yamamoto}}}]{Oka13}
\bibinfo{author}{\bibfnamefont{A.}~\bibnamefont{{Oka}}},
  \bibinfo{author}{\bibfnamefont{S.}~\bibnamefont{{Saito}}},
  \bibinfo{author}{\bibfnamefont{T.}~\bibnamefont{{Nishimichi}}},
  \bibinfo{author}{\bibfnamefont{A.}~\bibnamefont{{Taruya}}}, \bibnamefont{and}
  \bibinfo{author}{\bibfnamefont{K.}~\bibnamefont{{Yamamoto}}},
  \bibinfo{journal}{\mnras} \textbf{\bibinfo{volume}{439}},
  \bibinfo{pages}{2515} (\bibinfo{year}{2014}), \eprint{1310.2820}.

\bibitem[{\citenamefont{{Hikage} and {Yamamoto}}(2013)}]{HY2013}
\bibinfo{author}{\bibfnamefont{C.}~\bibnamefont{{Hikage}}} \bibnamefont{and}
  \bibinfo{author}{\bibfnamefont{K.}~\bibnamefont{{Yamamoto}}},
  \bibinfo{journal}{\jcap} \textbf{\bibinfo{volume}{8}}, \bibinfo{eid}{019}
  (\bibinfo{year}{2013}), \eprint{1303.3380}.

\bibitem[{\citenamefont{{Ivanov} et~al.}(2020)\citenamefont{{Ivanov},
  {Simonovi{\'c}}, and {Zaldarriaga}}}]{Ivanov20}
\bibinfo{author}{\bibfnamefont{M.~M.} \bibnamefont{{Ivanov}}},
  \bibinfo{author}{\bibfnamefont{M.}~\bibnamefont{{Simonovi{\'c}}}},
  \bibnamefont{and}
  \bibinfo{author}{\bibfnamefont{M.}~\bibnamefont{{Zaldarriaga}}},
  \bibinfo{journal}{\jcap} \textbf{\bibinfo{volume}{2020}}, \bibinfo{eid}{042}
  (\bibinfo{year}{2020}), \eprint{1909.05277}.

\bibitem[{\citenamefont{{Takada} et~al.}(2012)\citenamefont{{Takada}, {Ellis},
  {Chiba}, {Greene}, {Aihara}, {Arimoto}, {Bundy}, {Cohen}, {Dor{\'e}},
  {Graves} et~al.}}]{PFS}
\bibinfo{author}{\bibfnamefont{M.}~\bibnamefont{{Takada}}},
  \bibinfo{author}{\bibfnamefont{R.}~\bibnamefont{{Ellis}}},
  \bibinfo{author}{\bibfnamefont{M.}~\bibnamefont{{Chiba}}},
  \bibinfo{author}{\bibfnamefont{J.~E.} \bibnamefont{{Greene}}},
  \bibinfo{author}{\bibfnamefont{H.}~\bibnamefont{{Aihara}}},
  \bibinfo{author}{\bibfnamefont{N.}~\bibnamefont{{Arimoto}}},
  \bibinfo{author}{\bibfnamefont{K.}~\bibnamefont{{Bundy}}},
  \bibinfo{author}{\bibfnamefont{J.}~\bibnamefont{{Cohen}}},
  \bibinfo{author}{\bibfnamefont{O.}~\bibnamefont{{Dor{\'e}}}},
  \bibinfo{author}{\bibfnamefont{G.}~\bibnamefont{{Graves}}},
  \bibnamefont{et~al.}, \bibinfo{journal}{ArXiv e-prints}
  (\bibinfo{year}{2012}), \eprint{1206.0737}.

\bibitem[{\citenamefont{{DESI Collaboration} et~al.}(2016)\citenamefont{{DESI
  Collaboration}, {Aghamousa}, {Aguilar}, {Ahlen}, {Alam}, {Allen}, {Allende
  Prieto}, {Annis}, {Bailey}, {Balland} et~al.}}]{DESI16}
\bibinfo{author}{\bibnamefont{{DESI Collaboration}}},
  \bibinfo{author}{\bibfnamefont{A.}~\bibnamefont{{Aghamousa}}},
  \bibinfo{author}{\bibfnamefont{J.}~\bibnamefont{{Aguilar}}},
  \bibinfo{author}{\bibfnamefont{S.}~\bibnamefont{{Ahlen}}},
  \bibinfo{author}{\bibfnamefont{S.}~\bibnamefont{{Alam}}},
  \bibinfo{author}{\bibfnamefont{L.~E.} \bibnamefont{{Allen}}},
  \bibinfo{author}{\bibfnamefont{C.}~\bibnamefont{{Allende Prieto}}},
  \bibinfo{author}{\bibfnamefont{J.}~\bibnamefont{{Annis}}},
  \bibinfo{author}{\bibfnamefont{S.}~\bibnamefont{{Bailey}}},
  \bibinfo{author}{\bibfnamefont{C.}~\bibnamefont{{Balland}}},
  \bibnamefont{et~al.}, \bibinfo{journal}{ArXiv e-prints}
  (\bibinfo{year}{2016}), \eprint{1611.00036}.

\bibitem[{\citenamefont{{Hill} et~al.}(2008)\citenamefont{{Hill}, {Gebhardt},
  {Komatsu}, {Drory}, {MacQueen}, {Adams}, {Blanc}, {Koehler}, {Rafal}, {Roth}
  et~al.}}]{HETDEX}
\bibinfo{author}{\bibfnamefont{G.~J.} \bibnamefont{{Hill}}},
  \bibinfo{author}{\bibfnamefont{K.}~\bibnamefont{{Gebhardt}}},
  \bibinfo{author}{\bibfnamefont{E.}~\bibnamefont{{Komatsu}}},
  \bibinfo{author}{\bibfnamefont{N.}~\bibnamefont{{Drory}}},
  \bibinfo{author}{\bibfnamefont{P.~J.} \bibnamefont{{MacQueen}}},
  \bibinfo{author}{\bibfnamefont{J.}~\bibnamefont{{Adams}}},
  \bibinfo{author}{\bibfnamefont{G.~A.} \bibnamefont{{Blanc}}},
  \bibinfo{author}{\bibfnamefont{R.}~\bibnamefont{{Koehler}}},
  \bibinfo{author}{\bibfnamefont{M.}~\bibnamefont{{Rafal}}},
  \bibinfo{author}{\bibfnamefont{M.~M.} \bibnamefont{{Roth}}},
  \bibnamefont{et~al.}, in \emph{\bibinfo{booktitle}{Panoramic Views of Galaxy
  Formation and Evolution}}, edited by
  \bibinfo{editor}{\bibfnamefont{T.}~\bibnamefont{{Kodama}}},
  \bibinfo{editor}{\bibfnamefont{T.}~\bibnamefont{{Yamada}}}, \bibnamefont{and}
  \bibinfo{editor}{\bibfnamefont{K.}~\bibnamefont{{Aoki}}}
  (\bibinfo{year}{2008}), vol. \bibinfo{volume}{399} of
  \emph{\bibinfo{series}{Astronomical Society of the Pacific Conference
  Series}}, p. \bibinfo{pages}{115}, \eprint{0806.0183}.

\bibitem[{\citenamefont{{Amendola} et~al.}(2016)\citenamefont{{Amendola},
  {Appleby}, {Avgoustidis}, {Bacon}, {Baker}, {Baldi}, {Bartolo}, {Blanchard},
  {Bonvin}, {Borgani} et~al.}}]{Euclid16}
\bibinfo{author}{\bibfnamefont{L.}~\bibnamefont{{Amendola}}},
  \bibinfo{author}{\bibfnamefont{S.}~\bibnamefont{{Appleby}}},
  \bibinfo{author}{\bibfnamefont{A.}~\bibnamefont{{Avgoustidis}}},
  \bibinfo{author}{\bibfnamefont{D.}~\bibnamefont{{Bacon}}},
  \bibinfo{author}{\bibfnamefont{T.}~\bibnamefont{{Baker}}},
  \bibinfo{author}{\bibfnamefont{M.}~\bibnamefont{{Baldi}}},
  \bibinfo{author}{\bibfnamefont{N.}~\bibnamefont{{Bartolo}}},
  \bibinfo{author}{\bibfnamefont{A.}~\bibnamefont{{Blanchard}}},
  \bibinfo{author}{\bibfnamefont{C.}~\bibnamefont{{Bonvin}}},
  \bibinfo{author}{\bibfnamefont{S.}~\bibnamefont{{Borgani}}},
  \bibnamefont{et~al.}, \bibinfo{journal}{ArXiv e-prints}
  (\bibinfo{year}{2016}), \eprint{1606.00180}.

\bibitem[{\citenamefont{{Spergel} et~al.}(2015)\citenamefont{{Spergel},
  {Gehrels}, {Baltay}, {Bennett}, {Breckinridge}, {Donahue}, {Dressler},
  {Gaudi}, {Greene}, {Guyon} et~al.}}]{WFIRST15}
\bibinfo{author}{\bibfnamefont{D.}~\bibnamefont{{Spergel}}},
  \bibinfo{author}{\bibfnamefont{N.}~\bibnamefont{{Gehrels}}},
  \bibinfo{author}{\bibfnamefont{C.}~\bibnamefont{{Baltay}}},
  \bibinfo{author}{\bibfnamefont{D.}~\bibnamefont{{Bennett}}},
  \bibinfo{author}{\bibfnamefont{J.}~\bibnamefont{{Breckinridge}}},
  \bibinfo{author}{\bibfnamefont{M.}~\bibnamefont{{Donahue}}},
  \bibinfo{author}{\bibfnamefont{A.}~\bibnamefont{{Dressler}}},
  \bibinfo{author}{\bibfnamefont{B.~S.} \bibnamefont{{Gaudi}}},
  \bibinfo{author}{\bibfnamefont{T.}~\bibnamefont{{Greene}}},
  \bibinfo{author}{\bibfnamefont{O.}~\bibnamefont{{Guyon}}},
  \bibnamefont{et~al.}, \bibinfo{journal}{ArXiv e-prints}
  (\bibinfo{year}{2015}), \eprint{1503.03757}.

\bibitem[{\citenamefont{{Crocce} and
  {Scoccimarro}}(2008)}]{CrocceScoccimarro08}
\bibinfo{author}{\bibfnamefont{M.}~\bibnamefont{{Crocce}}} \bibnamefont{and}
  \bibinfo{author}{\bibfnamefont{R.}~\bibnamefont{{Scoccimarro}}},
  \bibinfo{journal}{\prd} \textbf{\bibinfo{volume}{77}}, \bibinfo{eid}{023533}
  (\bibinfo{year}{2008}), \eprint{0704.2783}.

\bibitem[{\citenamefont{{Scoccimarro}
  et~al.}(1999{\natexlab{a}})\citenamefont{{Scoccimarro}, {Zaldarriaga}, and
  {Hui}}}]{Scoccimarro99b}
\bibinfo{author}{\bibfnamefont{R.}~\bibnamefont{{Scoccimarro}}},
  \bibinfo{author}{\bibfnamefont{M.}~\bibnamefont{{Zaldarriaga}}},
  \bibnamefont{and} \bibinfo{author}{\bibfnamefont{L.}~\bibnamefont{{Hui}}},
  \bibinfo{journal}{\apj} \textbf{\bibinfo{volume}{527}}, \bibinfo{pages}{1}
  (\bibinfo{year}{1999}{\natexlab{a}}), \eprint{astro-ph/9901099}.

\bibitem[{\citenamefont{{Takahashi} et~al.}(2009)\citenamefont{{Takahashi},
  {Yoshida}, {Takada}, {Matsubara}, {Sugiyama}, {Kayo}, {Nishizawa},
  {Nishimichi}, {Saito}, and {Taruya}}}]{Takahashi09}
\bibinfo{author}{\bibfnamefont{R.}~\bibnamefont{{Takahashi}}},
  \bibinfo{author}{\bibfnamefont{N.}~\bibnamefont{{Yoshida}}},
  \bibinfo{author}{\bibfnamefont{M.}~\bibnamefont{{Takada}}},
  \bibinfo{author}{\bibfnamefont{T.}~\bibnamefont{{Matsubara}}},
  \bibinfo{author}{\bibfnamefont{N.}~\bibnamefont{{Sugiyama}}},
  \bibinfo{author}{\bibfnamefont{I.}~\bibnamefont{{Kayo}}},
  \bibinfo{author}{\bibfnamefont{A.~J.} \bibnamefont{{Nishizawa}}},
  \bibinfo{author}{\bibfnamefont{T.}~\bibnamefont{{Nishimichi}}},
  \bibinfo{author}{\bibfnamefont{S.}~\bibnamefont{{Saito}}}, \bibnamefont{and}
  \bibinfo{author}{\bibfnamefont{A.}~\bibnamefont{{Taruya}}},
  \bibinfo{journal}{\apj} \textbf{\bibinfo{volume}{700}}, \bibinfo{pages}{479}
  (\bibinfo{year}{2009}), \eprint{0902.0371}.

\bibitem[{\citenamefont{{Carron} et~al.}(2015)\citenamefont{{Carron}, {Wolk},
  and {Szapudi}}}]{Carron15}
\bibinfo{author}{\bibfnamefont{J.}~\bibnamefont{{Carron}}},
  \bibinfo{author}{\bibfnamefont{M.}~\bibnamefont{{Wolk}}}, \bibnamefont{and}
  \bibinfo{author}{\bibfnamefont{I.}~\bibnamefont{{Szapudi}}},
  \bibinfo{journal}{\mnras} \textbf{\bibinfo{volume}{453}},
  \bibinfo{pages}{450} (\bibinfo{year}{2015}), \eprint{1412.5511}.

\bibitem[{\citenamefont{{Eisenstein}
  et~al.}(2007{\natexlab{b}})\citenamefont{{Eisenstein}, {Seo}, {Sirko}, and
  {Spergel}}}]{Eisenstein07b}
\bibinfo{author}{\bibfnamefont{D.~J.} \bibnamefont{{Eisenstein}}},
  \bibinfo{author}{\bibfnamefont{H.-J.} \bibnamefont{{Seo}}},
  \bibinfo{author}{\bibfnamefont{E.}~\bibnamefont{{Sirko}}}, \bibnamefont{and}
  \bibinfo{author}{\bibfnamefont{D.~N.} \bibnamefont{{Spergel}}},
  \bibinfo{journal}{\apj} \textbf{\bibinfo{volume}{664}}, \bibinfo{pages}{675}
  (\bibinfo{year}{2007}{\natexlab{b}}), \eprint{astro-ph/0604362}.

\bibitem[{\citenamefont{{Zel'dovich}}(1970)}]{Zeldovich70}
\bibinfo{author}{\bibfnamefont{Y.~B.} \bibnamefont{{Zel'dovich}}},
  \bibinfo{journal}{\aap} \textbf{\bibinfo{volume}{5}}, \bibinfo{pages}{84}
  (\bibinfo{year}{1970}).

\bibitem[{\citenamefont{{Seo} et~al.}(2008)\citenamefont{{Seo}, {Siegel},
  {Eisenstein}, and {White}}}]{Seo08}
\bibinfo{author}{\bibfnamefont{H.-J.} \bibnamefont{{Seo}}},
  \bibinfo{author}{\bibfnamefont{E.~R.} \bibnamefont{{Siegel}}},
  \bibinfo{author}{\bibfnamefont{D.~J.} \bibnamefont{{Eisenstein}}},
  \bibnamefont{and} \bibinfo{author}{\bibfnamefont{M.}~\bibnamefont{{White}}},
  \bibinfo{journal}{\apj} \textbf{\bibinfo{volume}{686}}, \bibinfo{pages}{13}
  (\bibinfo{year}{2008}), \eprint{0805.0117}.

\bibitem[{\citenamefont{{Padmanabhan} et~al.}(2009)\citenamefont{{Padmanabhan},
  {White}, and {Cohn}}}]{Padmanabhan09}
\bibinfo{author}{\bibfnamefont{N.}~\bibnamefont{{Padmanabhan}}},
  \bibinfo{author}{\bibfnamefont{M.}~\bibnamefont{{White}}}, \bibnamefont{and}
  \bibinfo{author}{\bibfnamefont{J.~D.} \bibnamefont{{Cohn}}},
  \bibinfo{journal}{\prd} \textbf{\bibinfo{volume}{79}}, \bibinfo{eid}{063523}
  (\bibinfo{year}{2009}), \eprint{0812.2905}.

\bibitem[{\citenamefont{{Noh} et~al.}(2009)\citenamefont{{Noh}, {White}, and
  {Padmanabhan}}}]{Noh09}
\bibinfo{author}{\bibfnamefont{Y.}~\bibnamefont{{Noh}}},
  \bibinfo{author}{\bibfnamefont{M.}~\bibnamefont{{White}}}, \bibnamefont{and}
  \bibinfo{author}{\bibfnamefont{N.}~\bibnamefont{{Padmanabhan}}},
  \bibinfo{journal}{\prd} \textbf{\bibinfo{volume}{80}}, \bibinfo{eid}{123501}
  (\bibinfo{year}{2009}), \eprint{0909.1802}.

\bibitem[{\citenamefont{{Seo} et~al.}(2010)\citenamefont{{Seo}, {Eckel},
  {Eisenstein}, {Mehta}, {Metchnik}, {Padmanabhan}, {Pinto}, {Takahashi},
  {White}, and {Xu}}}]{Seo10}
\bibinfo{author}{\bibfnamefont{H.-J.} \bibnamefont{{Seo}}},
  \bibinfo{author}{\bibfnamefont{J.}~\bibnamefont{{Eckel}}},
  \bibinfo{author}{\bibfnamefont{D.~J.} \bibnamefont{{Eisenstein}}},
  \bibinfo{author}{\bibfnamefont{K.}~\bibnamefont{{Mehta}}},
  \bibinfo{author}{\bibfnamefont{M.}~\bibnamefont{{Metchnik}}},
  \bibinfo{author}{\bibfnamefont{N.}~\bibnamefont{{Padmanabhan}}},
  \bibinfo{author}{\bibfnamefont{P.}~\bibnamefont{{Pinto}}},
  \bibinfo{author}{\bibfnamefont{R.}~\bibnamefont{{Takahashi}}},
  \bibinfo{author}{\bibfnamefont{M.}~\bibnamefont{{White}}}, \bibnamefont{and}
  \bibinfo{author}{\bibfnamefont{X.}~\bibnamefont{{Xu}}},
  \bibinfo{journal}{\apj} \textbf{\bibinfo{volume}{720}}, \bibinfo{pages}{1650}
  (\bibinfo{year}{2010}), \eprint{0910.5005}.

\bibitem[{\citenamefont{{Sherwin} and {Zaldarriaga}}(2012)}]{Sherwin12}
\bibinfo{author}{\bibfnamefont{B.~D.} \bibnamefont{{Sherwin}}}
  \bibnamefont{and}
  \bibinfo{author}{\bibfnamefont{M.}~\bibnamefont{{Zaldarriaga}}},
  \bibinfo{journal}{\prd} \textbf{\bibinfo{volume}{85}}, \bibinfo{eid}{103523}
  (\bibinfo{year}{2012}), \eprint{1202.3998}.

\bibitem[{\citenamefont{{Padmanabhan} et~al.}(2012)\citenamefont{{Padmanabhan},
  {Xu}, {Eisenstein}, {Scalzo}, {Cuesta}, {Mehta}, and
  {Kazin}}}]{Padmanabhan12}
\bibinfo{author}{\bibfnamefont{N.}~\bibnamefont{{Padmanabhan}}},
  \bibinfo{author}{\bibfnamefont{X.}~\bibnamefont{{Xu}}},
  \bibinfo{author}{\bibfnamefont{D.~J.} \bibnamefont{{Eisenstein}}},
  \bibinfo{author}{\bibfnamefont{R.}~\bibnamefont{{Scalzo}}},
  \bibinfo{author}{\bibfnamefont{A.~J.} \bibnamefont{{Cuesta}}},
  \bibinfo{author}{\bibfnamefont{K.~T.} \bibnamefont{{Mehta}}},
  \bibnamefont{and} \bibinfo{author}{\bibfnamefont{E.}~\bibnamefont{{Kazin}}},
  \bibinfo{journal}{\mnras} \textbf{\bibinfo{volume}{427}},
  \bibinfo{pages}{2132} (\bibinfo{year}{2012}), \eprint{1202.0090}.

\bibitem[{\citenamefont{{Tassev} and
  {Zaldarriaga}}(2012)}]{TassevZaldarriaga12}
\bibinfo{author}{\bibfnamefont{S.}~\bibnamefont{{Tassev}}} \bibnamefont{and}
  \bibinfo{author}{\bibfnamefont{M.}~\bibnamefont{{Zaldarriaga}}},
  \bibinfo{journal}{\jcap} \textbf{\bibinfo{volume}{2012}}, \bibinfo{eid}{006}
  (\bibinfo{year}{2012}), \eprint{1203.6066}.

\bibitem[{\citenamefont{{Schmittfull} et~al.}(2015)\citenamefont{{Schmittfull},
  {Feng}, {Beutler}, {Sherwin}, and {Chu}}}]{Schmittfull15}
\bibinfo{author}{\bibfnamefont{M.}~\bibnamefont{{Schmittfull}}},
  \bibinfo{author}{\bibfnamefont{Y.}~\bibnamefont{{Feng}}},
  \bibinfo{author}{\bibfnamefont{F.}~\bibnamefont{{Beutler}}},
  \bibinfo{author}{\bibfnamefont{B.}~\bibnamefont{{Sherwin}}},
  \bibnamefont{and} \bibinfo{author}{\bibfnamefont{M.~Y.} \bibnamefont{{Chu}}},
  \bibinfo{journal}{\prd} \textbf{\bibinfo{volume}{92}}, \bibinfo{eid}{123522}
  (\bibinfo{year}{2015}), \eprint{1508.06972}.

\bibitem[{\citenamefont{{Seo} et~al.}(2016)\citenamefont{{Seo}, {Beutler},
  {Ross}, and {Saito}}}]{Seo16}
\bibinfo{author}{\bibfnamefont{H.-J.} \bibnamefont{{Seo}}},
  \bibinfo{author}{\bibfnamefont{F.}~\bibnamefont{{Beutler}}},
  \bibinfo{author}{\bibfnamefont{A.~J.} \bibnamefont{{Ross}}},
  \bibnamefont{and} \bibinfo{author}{\bibfnamefont{S.}~\bibnamefont{{Saito}}},
  \bibinfo{journal}{\mnras} \textbf{\bibinfo{volume}{460}},
  \bibinfo{pages}{2453} (\bibinfo{year}{2016}), \eprint{1511.00663}.

\bibitem[{\citenamefont{{Schmittfull} et~al.}(2017)\citenamefont{{Schmittfull},
  {Baldauf}, and {Zaldarriaga}}}]{Schmittfull17}
\bibinfo{author}{\bibfnamefont{M.}~\bibnamefont{{Schmittfull}}},
  \bibinfo{author}{\bibfnamefont{T.}~\bibnamefont{{Baldauf}}},
  \bibnamefont{and}
  \bibinfo{author}{\bibfnamefont{M.}~\bibnamefont{{Zaldarriaga}}},
  \bibinfo{journal}{\prd} \textbf{\bibinfo{volume}{96}}, \bibinfo{eid}{023505}
  (\bibinfo{year}{2017}), \eprint{1704.06634}.

\bibitem[{\citenamefont{{Wang} et~al.}(2017)\citenamefont{{Wang}, {Yu}, {Zhu},
  {Yu}, {Pan}, and {Pen}}}]{Wang17}
\bibinfo{author}{\bibfnamefont{X.}~\bibnamefont{{Wang}}},
  \bibinfo{author}{\bibfnamefont{H.-R.} \bibnamefont{{Yu}}},
  \bibinfo{author}{\bibfnamefont{H.-M.} \bibnamefont{{Zhu}}},
  \bibinfo{author}{\bibfnamefont{Y.}~\bibnamefont{{Yu}}},
  \bibinfo{author}{\bibfnamefont{Q.}~\bibnamefont{{Pan}}}, \bibnamefont{and}
  \bibinfo{author}{\bibfnamefont{U.-L.} \bibnamefont{{Pen}}},
  \bibinfo{journal}{\apjl} \textbf{\bibinfo{volume}{841}}, \bibinfo{eid}{L29}
  (\bibinfo{year}{2017}), \eprint{1703.09742}.

\bibitem[{\citenamefont{{Yu} et~al.}(2017)\citenamefont{{Yu}, {Zhu}, and
  {Pen}}}]{Yu17}
\bibinfo{author}{\bibfnamefont{Y.}~\bibnamefont{{Yu}}},
  \bibinfo{author}{\bibfnamefont{H.-M.} \bibnamefont{{Zhu}}}, \bibnamefont{and}
  \bibinfo{author}{\bibfnamefont{U.-L.} \bibnamefont{{Pen}}},
  \bibinfo{journal}{\apj} \textbf{\bibinfo{volume}{847}}, \bibinfo{eid}{110}
  (\bibinfo{year}{2017}), \eprint{1703.08301}.

\bibitem[{\citenamefont{{Hada} and {Eisenstein}}(2018)}]{Hada18}
\bibinfo{author}{\bibfnamefont{R.}~\bibnamefont{{Hada}}} \bibnamefont{and}
  \bibinfo{author}{\bibfnamefont{D.~J.} \bibnamefont{{Eisenstein}}},
  \bibinfo{journal}{\mnras} \textbf{\bibinfo{volume}{478}},
  \bibinfo{pages}{1866} (\bibinfo{year}{2018}), \eprint{1804.04738}.

\bibitem[{\citenamefont{{Hikage} et~al.}(2017)\citenamefont{{Hikage}, {Koyama},
  and {Heavens}}}]{HKH17}
\bibinfo{author}{\bibfnamefont{C.}~\bibnamefont{{Hikage}}},
  \bibinfo{author}{\bibfnamefont{K.}~\bibnamefont{{Koyama}}}, \bibnamefont{and}
  \bibinfo{author}{\bibfnamefont{A.}~\bibnamefont{{Heavens}}},
  \bibinfo{journal}{\prd} \textbf{\bibinfo{volume}{96}}, \bibinfo{eid}{043513}
  (\bibinfo{year}{2017}), \eprint{1703.07878}.

\bibitem[{\citenamefont{{Hikage} et~al.}(2020)\citenamefont{{Hikage}, {Koyama},
  and {Takahashi}}}]{HKT19}
\bibinfo{author}{\bibfnamefont{C.}~\bibnamefont{{Hikage}}},
  \bibinfo{author}{\bibfnamefont{K.}~\bibnamefont{{Koyama}}}, \bibnamefont{and}
  \bibinfo{author}{\bibfnamefont{R.}~\bibnamefont{{Takahashi}}},
  \bibinfo{journal}{\prd} \textbf{\bibinfo{volume}{101}}, \bibinfo{eid}{043510}
  (\bibinfo{year}{2020}), \eprint{1911.06461}.

\bibitem[{\citenamefont{{Harnois-D{\'e}raps}
  et~al.}(2013)\citenamefont{{Harnois-D{\'e}raps}, {Pen}, {Iliev}, {Merz},
  {Emberson}, and {Desjacques}}}]{HarnoisDeraps13}
\bibinfo{author}{\bibfnamefont{J.}~\bibnamefont{{Harnois-D{\'e}raps}}},
  \bibinfo{author}{\bibfnamefont{U.-L.} \bibnamefont{{Pen}}},
  \bibinfo{author}{\bibfnamefont{I.~T.} \bibnamefont{{Iliev}}},
  \bibinfo{author}{\bibfnamefont{H.}~\bibnamefont{{Merz}}},
  \bibinfo{author}{\bibfnamefont{J.~D.} \bibnamefont{{Emberson}}},
  \bibnamefont{and}
  \bibinfo{author}{\bibfnamefont{V.}~\bibnamefont{{Desjacques}}},
  \bibinfo{journal}{\mnras} \textbf{\bibinfo{volume}{436}},
  \bibinfo{pages}{540} (\bibinfo{year}{2013}), \eprint{1208.5098}.

\bibitem[{\citenamefont{{Blot} et~al.}(2015)\citenamefont{{Blot}, {Corasaniti},
  {Alimi}, {Reverdy}, and {Rasera}}}]{Blot15}
\bibinfo{author}{\bibfnamefont{L.}~\bibnamefont{{Blot}}},
  \bibinfo{author}{\bibfnamefont{P.~S.} \bibnamefont{{Corasaniti}}},
  \bibinfo{author}{\bibfnamefont{J.~M.} \bibnamefont{{Alimi}}},
  \bibinfo{author}{\bibfnamefont{V.}~\bibnamefont{{Reverdy}}},
  \bibnamefont{and} \bibinfo{author}{\bibfnamefont{Y.}~\bibnamefont{{Rasera}}},
  \bibinfo{journal}{\mnras} \textbf{\bibinfo{volume}{446}},
  \bibinfo{pages}{1756} (\bibinfo{year}{2015}), \eprint{1406.2713}.

\bibitem[{\citenamefont{{Klypin} and {Prada}}(2018)}]{Klypin18}
\bibinfo{author}{\bibfnamefont{A.}~\bibnamefont{{Klypin}}} \bibnamefont{and}
  \bibinfo{author}{\bibfnamefont{F.}~\bibnamefont{{Prada}}},
  \bibinfo{journal}{\mnras} \textbf{\bibinfo{volume}{478}},
  \bibinfo{pages}{4602} (\bibinfo{year}{2018}), \eprint{1701.05690}.

\bibitem[{\citenamefont{{Villaescusa-Navarro}
  et~al.}(2019)\citenamefont{{Villaescusa-Navarro}, {Hahn}, {Massara},
  {Banerjee}, {Delgado}, {Kodi Ramanah}, {Charnock}, {Giusarma}, {Li}, {Allys}
  et~al.}}]{Quijote19}
\bibinfo{author}{\bibfnamefont{F.}~\bibnamefont{{Villaescusa-Navarro}}},
  \bibinfo{author}{\bibfnamefont{C.}~\bibnamefont{{Hahn}}},
  \bibinfo{author}{\bibfnamefont{E.}~\bibnamefont{{Massara}}},
  \bibinfo{author}{\bibfnamefont{A.}~\bibnamefont{{Banerjee}}},
  \bibinfo{author}{\bibfnamefont{A.~M.} \bibnamefont{{Delgado}}},
  \bibinfo{author}{\bibfnamefont{D.}~\bibnamefont{{Kodi Ramanah}}},
  \bibinfo{author}{\bibfnamefont{T.}~\bibnamefont{{Charnock}}},
  \bibinfo{author}{\bibfnamefont{E.}~\bibnamefont{{Giusarma}}},
  \bibinfo{author}{\bibfnamefont{Y.}~\bibnamefont{{Li}}},
  \bibinfo{author}{\bibfnamefont{E.}~\bibnamefont{{Allys}}},
  \bibnamefont{et~al.}, \bibinfo{journal}{arXiv e-prints}
  \bibinfo{eid}{arXiv:1909.05273} (\bibinfo{year}{2019}), \eprint{1909.05273}.

\bibitem[{\citenamefont{{Hamilton} et~al.}(2006)\citenamefont{{Hamilton},
  {Rimes}, and {Scoccimarro}}}]{Hamilton06}
\bibinfo{author}{\bibfnamefont{A.~J.~S.} \bibnamefont{{Hamilton}}},
  \bibinfo{author}{\bibfnamefont{C.~D.} \bibnamefont{{Rimes}}},
  \bibnamefont{and}
  \bibinfo{author}{\bibfnamefont{R.}~\bibnamefont{{Scoccimarro}}},
  \bibinfo{journal}{\mnras} \textbf{\bibinfo{volume}{371}},
  \bibinfo{pages}{1188} (\bibinfo{year}{2006}), \eprint{astro-ph/0511416}.

\bibitem[{\citenamefont{{Takada} and {Hu}}(2013)}]{TakadaHu13}
\bibinfo{author}{\bibfnamefont{M.}~\bibnamefont{{Takada}}} \bibnamefont{and}
  \bibinfo{author}{\bibfnamefont{W.}~\bibnamefont{{Hu}}},
  \bibinfo{journal}{\prd} \textbf{\bibinfo{volume}{87}}, \bibinfo{eid}{123504}
  (\bibinfo{year}{2013}), \eprint{1302.6994}.

\bibitem[{\citenamefont{{Li} et~al.}(2014)\citenamefont{{Li}, {Hu}, and
  {Takada}}}]{Li14}
\bibinfo{author}{\bibfnamefont{Y.}~\bibnamefont{{Li}}},
  \bibinfo{author}{\bibfnamefont{W.}~\bibnamefont{{Hu}}}, \bibnamefont{and}
  \bibinfo{author}{\bibfnamefont{M.}~\bibnamefont{{Takada}}},
  \bibinfo{journal}{\prd} \textbf{\bibinfo{volume}{89}}, \bibinfo{eid}{083519}
  (\bibinfo{year}{2014}), \eprint{1401.0385}.

\bibitem[{\citenamefont{{Planck Collaboration}
  et~al.}(2016)\citenamefont{{Planck Collaboration}, {Ade}, {Aghanim},
  {Arnaud}, {Ashdown}, {Aumont}, {Baccigalupi}, {Banday}, {Barreiro},
  {Bartlett} et~al.}}]{Planck15}
\bibinfo{author}{\bibnamefont{{Planck Collaboration}}},
  \bibinfo{author}{\bibfnamefont{P.~A.~R.} \bibnamefont{{Ade}}},
  \bibinfo{author}{\bibfnamefont{N.}~\bibnamefont{{Aghanim}}},
  \bibinfo{author}{\bibfnamefont{M.}~\bibnamefont{{Arnaud}}},
  \bibinfo{author}{\bibfnamefont{M.}~\bibnamefont{{Ashdown}}},
  \bibinfo{author}{\bibfnamefont{J.}~\bibnamefont{{Aumont}}},
  \bibinfo{author}{\bibfnamefont{C.}~\bibnamefont{{Baccigalupi}}},
  \bibinfo{author}{\bibfnamefont{A.~J.} \bibnamefont{{Banday}}},
  \bibinfo{author}{\bibfnamefont{R.~B.} \bibnamefont{{Barreiro}}},
  \bibinfo{author}{\bibfnamefont{J.~G.} \bibnamefont{{Bartlett}}},
  \bibnamefont{et~al.}, \bibinfo{journal}{\aap} \textbf{\bibinfo{volume}{594}},
  \bibinfo{eid}{A13} (\bibinfo{year}{2016}), \eprint{1502.01589}.

\bibitem[{\citenamefont{{Fry}}(1984)}]{Fry84}
\bibinfo{author}{\bibfnamefont{J.~N.} \bibnamefont{{Fry}}},
  \bibinfo{journal}{\apj} \textbf{\bibinfo{volume}{279}}, \bibinfo{pages}{499}
  (\bibinfo{year}{1984}).

\bibitem[{\citenamefont{{Scoccimarro}
  et~al.}(1999{\natexlab{b}})\citenamefont{{Scoccimarro}, {Couchman}, and
  {Frieman}}}]{Scoccimarro99a}
\bibinfo{author}{\bibfnamefont{R.}~\bibnamefont{{Scoccimarro}}},
  \bibinfo{author}{\bibfnamefont{H.~M.~P.} \bibnamefont{{Couchman}}},
  \bibnamefont{and} \bibinfo{author}{\bibfnamefont{J.~A.}
  \bibnamefont{{Frieman}}}, \bibinfo{journal}{\apj}
  \textbf{\bibinfo{volume}{517}}, \bibinfo{pages}{531}
  (\bibinfo{year}{1999}{\natexlab{b}}), \eprint{astro-ph/9808305}.

\bibitem[{\citenamefont{{Matsubara}}(2008)}]{Matsubara08a}
\bibinfo{author}{\bibfnamefont{T.}~\bibnamefont{{Matsubara}}},
  \bibinfo{journal}{\prd} \textbf{\bibinfo{volume}{77}}, \bibinfo{eid}{063530}
  (\bibinfo{year}{2008}), \eprint{0711.2521}.

\bibitem[{\citenamefont{{Wadekar} and {Scoccimarro}}(2019)}]{Wadekar19}
\bibinfo{author}{\bibfnamefont{D.}~\bibnamefont{{Wadekar}}} \bibnamefont{and}
  \bibinfo{author}{\bibfnamefont{R.}~\bibnamefont{{Scoccimarro}}},
  \bibinfo{journal}{arXiv e-prints} \bibinfo{eid}{arXiv:1910.02914}
  (\bibinfo{year}{2019}), \eprint{1910.02914}.

\bibitem[{\citenamefont{{Springel}}(2005)}]{Springel05}
\bibinfo{author}{\bibfnamefont{V.}~\bibnamefont{{Springel}}},
  \bibinfo{journal}{\mnras} \textbf{\bibinfo{volume}{364}},
  \bibinfo{pages}{1105} (\bibinfo{year}{2005}), \eprint{astro-ph/0505010}.

\bibitem[{\citenamefont{{Crocce} et~al.}(2006)\citenamefont{{Crocce},
  {Pueblas}, and {Scoccimarro}}}]{CPS2006}
\bibinfo{author}{\bibfnamefont{M.}~\bibnamefont{{Crocce}}},
  \bibinfo{author}{\bibfnamefont{S.}~\bibnamefont{{Pueblas}}},
  \bibnamefont{and}
  \bibinfo{author}{\bibfnamefont{R.}~\bibnamefont{{Scoccimarro}}},
  \bibinfo{journal}{\mnras} \textbf{\bibinfo{volume}{373}},
  \bibinfo{pages}{369} (\bibinfo{year}{2006}), \eprint{astro-ph/0606505}.

\bibitem[{\citenamefont{{Nishimichi} et~al.}(2009)\citenamefont{{Nishimichi},
  {Shirata}, {Taruya}, {Yahata}, {Saito}, {Suto}, {Takahashi}, {Yoshida},
  {Matsubara}, {Sugiyama} et~al.}}]{Nishimichi2009}
\bibinfo{author}{\bibfnamefont{T.}~\bibnamefont{{Nishimichi}}},
  \bibinfo{author}{\bibfnamefont{A.}~\bibnamefont{{Shirata}}},
  \bibinfo{author}{\bibfnamefont{A.}~\bibnamefont{{Taruya}}},
  \bibinfo{author}{\bibfnamefont{K.}~\bibnamefont{{Yahata}}},
  \bibinfo{author}{\bibfnamefont{S.}~\bibnamefont{{Saito}}},
  \bibinfo{author}{\bibfnamefont{Y.}~\bibnamefont{{Suto}}},
  \bibinfo{author}{\bibfnamefont{R.}~\bibnamefont{{Takahashi}}},
  \bibinfo{author}{\bibfnamefont{N.}~\bibnamefont{{Yoshida}}},
  \bibinfo{author}{\bibfnamefont{T.}~\bibnamefont{{Matsubara}}},
  \bibinfo{author}{\bibfnamefont{N.}~\bibnamefont{{Sugiyama}}},
  \bibnamefont{et~al.}, \bibinfo{journal}{\pasj} \textbf{\bibinfo{volume}{61}},
  \bibinfo{pages}{321} (\bibinfo{year}{2009}), \eprint{0810.0813}.

\bibitem[{\citenamefont{{Nishimichi} et~al.}(2019)\citenamefont{{Nishimichi},
  {Takada}, {Takahashi}, {Osato}, {Shirasaki}, {Oogi}, {Miyatake}, {Oguri},
  {Murata}, {Kobayashi} et~al.}}]{Nishimichi2019}
\bibinfo{author}{\bibfnamefont{T.}~\bibnamefont{{Nishimichi}}},
  \bibinfo{author}{\bibfnamefont{M.}~\bibnamefont{{Takada}}},
  \bibinfo{author}{\bibfnamefont{R.}~\bibnamefont{{Takahashi}}},
  \bibinfo{author}{\bibfnamefont{K.}~\bibnamefont{{Osato}}},
  \bibinfo{author}{\bibfnamefont{M.}~\bibnamefont{{Shirasaki}}},
  \bibinfo{author}{\bibfnamefont{T.}~\bibnamefont{{Oogi}}},
  \bibinfo{author}{\bibfnamefont{H.}~\bibnamefont{{Miyatake}}},
  \bibinfo{author}{\bibfnamefont{M.}~\bibnamefont{{Oguri}}},
  \bibinfo{author}{\bibfnamefont{R.}~\bibnamefont{{Murata}}},
  \bibinfo{author}{\bibfnamefont{Y.}~\bibnamefont{{Kobayashi}}},
  \bibnamefont{et~al.}, \bibinfo{journal}{\apj} \textbf{\bibinfo{volume}{884}},
  \bibinfo{eid}{29} (\bibinfo{year}{2019}), \eprint{1811.09504}.

\bibitem[{\citenamefont{{Lewis} et~al.}(2000)\citenamefont{{Lewis},
  {Challinor}, and {Lasenby}}}]{Lewis2000}
\bibinfo{author}{\bibfnamefont{A.}~\bibnamefont{{Lewis}}},
  \bibinfo{author}{\bibfnamefont{A.}~\bibnamefont{{Challinor}}},
  \bibnamefont{and}
  \bibinfo{author}{\bibfnamefont{A.}~\bibnamefont{{Lasenby}}},
  \bibinfo{journal}{\apj} \textbf{\bibinfo{volume}{538}}, \bibinfo{pages}{473}
  (\bibinfo{year}{2000}), \eprint{astro-ph/9911177}.

\bibitem[{Note1()}]{Note1}
Note1, \bibinfo{note}{fFTW3 at http://www.fftw.org}.

\bibitem[{\citenamefont{{Jing}}(2005)}]{Jing05}
\bibinfo{author}{\bibfnamefont{Y.~P.} \bibnamefont{{Jing}}},
  \bibinfo{journal}{\apj} \textbf{\bibinfo{volume}{620}}, \bibinfo{pages}{559}
  (\bibinfo{year}{2005}), \eprint{astro-ph/0409240}.

\bibitem[{\citenamefont{{Hartlap} et~al.}(2007)\citenamefont{{Hartlap},
  {Simon}, and {Schneider}}}]{Hartlap07}
\bibinfo{author}{\bibfnamefont{J.}~\bibnamefont{{Hartlap}}},
  \bibinfo{author}{\bibfnamefont{P.}~\bibnamefont{{Simon}}}, \bibnamefont{and}
  \bibinfo{author}{\bibfnamefont{P.}~\bibnamefont{{Schneider}}},
  \bibinfo{journal}{\aap} \textbf{\bibinfo{volume}{464}}, \bibinfo{pages}{399}
  (\bibinfo{year}{2007}), \eprint{astro-ph/0608064}.

\bibitem[{\citenamefont{{Carrasco} et~al.}(2012)\citenamefont{{Carrasco},
  {Hertzberg}, and {Senatore}}}]{Carrasco12}
\bibinfo{author}{\bibfnamefont{J.~J.~M.} \bibnamefont{{Carrasco}}},
  \bibinfo{author}{\bibfnamefont{M.~P.} \bibnamefont{{Hertzberg}}},
  \bibnamefont{and}
  \bibinfo{author}{\bibfnamefont{L.}~\bibnamefont{{Senatore}}},
  \bibinfo{journal}{Journal of High Energy Physics}
  \textbf{\bibinfo{volume}{2012}}, \bibinfo{eid}{82} (\bibinfo{year}{2012}),
  \eprint{1206.2926}.

\bibitem[{\citenamefont{{Feroz} et~al.}(2009)\citenamefont{{Feroz}, {Hobson},
  and {Bridges}}}]{Feroz09}
\bibinfo{author}{\bibfnamefont{F.}~\bibnamefont{{Feroz}}},
  \bibinfo{author}{\bibfnamefont{M.~P.} \bibnamefont{{Hobson}}},
  \bibnamefont{and}
  \bibinfo{author}{\bibfnamefont{M.}~\bibnamefont{{Bridges}}},
  \bibinfo{journal}{\mnras} \textbf{\bibinfo{volume}{398}},
  \bibinfo{pages}{1601} (\bibinfo{year}{2009}), \eprint{0809.3437}.

\bibitem[{\citenamefont{Audren et~al.}(2013)\citenamefont{Audren, Lesgourgues,
  Benabed, and Prunet}}]{Audren13}
\bibinfo{author}{\bibfnamefont{B.}~\bibnamefont{Audren}},
  \bibinfo{author}{\bibfnamefont{J.}~\bibnamefont{Lesgourgues}},
  \bibinfo{author}{\bibfnamefont{K.}~\bibnamefont{Benabed}}, \bibnamefont{and}
  \bibinfo{author}{\bibfnamefont{S.}~\bibnamefont{Prunet}},
  \bibinfo{journal}{Journal of Cosmology and Astroparticle Physics}
  \textbf{\bibinfo{volume}{2013}}, \bibinfo{pages}{001–001}
  (\bibinfo{year}{2013}), ISSN \bibinfo{issn}{1475-7516},
  \urlprefix\url{http://dx.doi.org/10.1088/1475-7516/2013/02/001}.

\bibitem[{\citenamefont{{Li} et~al.}(2018)\citenamefont{{Li}, {Schmittfull},
  and {Seljak}}}]{Li18}
\bibinfo{author}{\bibfnamefont{Y.}~\bibnamefont{{Li}}},
  \bibinfo{author}{\bibfnamefont{M.}~\bibnamefont{{Schmittfull}}},
  \bibnamefont{and} \bibinfo{author}{\bibfnamefont{U.}~\bibnamefont{{Seljak}}},
  \bibinfo{journal}{\jcap} \textbf{\bibinfo{volume}{2018}}, \bibinfo{eid}{022}
  (\bibinfo{year}{2018}), \eprint{1711.00018}.

\bibitem[{\citenamefont{{Bertolini} et~al.}(2016)\citenamefont{{Bertolini},
  {Schutz}, {Solon}, {Walsh}, and {Zurek}}}]{Bertolini16}
\bibinfo{author}{\bibfnamefont{D.}~\bibnamefont{{Bertolini}}},
  \bibinfo{author}{\bibfnamefont{K.}~\bibnamefont{{Schutz}}},
  \bibinfo{author}{\bibfnamefont{M.~P.} \bibnamefont{{Solon}}},
  \bibinfo{author}{\bibfnamefont{J.~R.} \bibnamefont{{Walsh}}},
  \bibnamefont{and} \bibinfo{author}{\bibfnamefont{K.~M.}
  \bibnamefont{{Zurek}}}, \bibinfo{journal}{\prd}
  \textbf{\bibinfo{volume}{93}}, \bibinfo{eid}{123505} (\bibinfo{year}{2016}),
  \eprint{1512.07630}.

\bibitem[{\citenamefont{{Barreira} and {Schmidt}}(2017)}]{Barreira17}
\bibinfo{author}{\bibfnamefont{A.}~\bibnamefont{{Barreira}}} \bibnamefont{and}
  \bibinfo{author}{\bibfnamefont{F.}~\bibnamefont{{Schmidt}}},
  \bibinfo{journal}{\jcap} \textbf{\bibinfo{volume}{2017}}, \bibinfo{eid}{051}
  (\bibinfo{year}{2017}), \eprint{1705.01092}.

\bibitem[{\citenamefont{{Neyrinck}}(2011)}]{Neyrinck11}
\bibinfo{author}{\bibfnamefont{M.~C.} \bibnamefont{{Neyrinck}}},
  \bibinfo{journal}{\apj} \textbf{\bibinfo{volume}{736}}, \bibinfo{eid}{8}
  (\bibinfo{year}{2011}), \eprint{1103.5476}.

\bibitem[{\citenamefont{{Mohammed} and {Seljak}}(2014)}]{Mohammed14}
\bibinfo{author}{\bibfnamefont{I.}~\bibnamefont{{Mohammed}}} \bibnamefont{and}
  \bibinfo{author}{\bibfnamefont{U.}~\bibnamefont{{Seljak}}},
  \bibinfo{journal}{\mnras} \textbf{\bibinfo{volume}{445}},
  \bibinfo{pages}{3382} (\bibinfo{year}{2014}), \eprint{1407.0060}.

\bibitem[{\citenamefont{{Mohammed} et~al.}(2017)\citenamefont{{Mohammed},
  {Seljak}, and {Vlah}}}]{Mohammed17}
\bibinfo{author}{\bibfnamefont{I.}~\bibnamefont{{Mohammed}}},
  \bibinfo{author}{\bibfnamefont{U.}~\bibnamefont{{Seljak}}}, \bibnamefont{and}
  \bibinfo{author}{\bibfnamefont{Z.}~\bibnamefont{{Vlah}}},
  \bibinfo{journal}{\mnras} \textbf{\bibinfo{volume}{466}},
  \bibinfo{pages}{780} (\bibinfo{year}{2017}), \eprint{1607.00043}.

\end{thebibliography}

\end{document}